\documentclass[pra, twocolumn,superscriptaddress]{revtex4}
\usepackage[utf8]{inputenc}
\usepackage[american,british]{babel}
\usepackage[T1]{fontenc}
\usepackage[pdftex]{graphicx}  
\usepackage{graphicx}
\usepackage{xcolor}
\usepackage{mathtools}
\usepackage{bbm}
\usepackage{subfigure}
\usepackage{dcolumn}
\usepackage{braket}
\usepackage{bm}
\usepackage{amsmath,amsthm,amssymb}
\usepackage{color}
\usepackage{verbatim}
\usepackage{comment}
\usepackage{accents}

\usepackage{appendix}

\definecolor{Red}{HTML}{E53E30}  
\definecolor{Green}{HTML}{00AD69}  
\definecolor{Blue}{HTML}{2171b5}
\definecolor{Purple}{HTML}{652F6C}  

\usepackage{bbold}

\usepackage[colorlinks=true,
   linkcolor=blue,
   citecolor=blue,
   urlcolor =blue]{hyperref}

\newcommand{\citeSM}{\cite[{\tiny SM}\kern-0.3em][]{SM}}
\definecolor{bluegray}{rgb}{0.2, 0.4, 0.6}

\newcommand{\be}{\begin{equation}}
\newcommand{\ee}{\end{equation}}

\newcommand{\Tr}{\mathrm{Tr}}

\usepackage{array}
\newcommand{\PreserveBackslash}[1]{\let\temp=\\#1\let\\=\temp}
\newcolumntype{C}[1]{>{\PreserveBackslash\centering}p{#1}}

\newcommand{\expp}{\mathbb{E}}

\DeclarePairedDelimiter\floor{\lfloor}{\rfloor}
\usepackage[normalem]{ulem}

\newcommand{\s}{\mathbf{s}}
\renewcommand{\S}{\mathbb{S}}
\newcommand{\0}{\mathbf{0}}

\usepackage{dsfont}
\usepackage{physics}

\begin{document}
\title{Robust estimation of the Quantum Fisher Information on a quantum processor}

\author{Vittorio Vitale}
\affiliation{Univ.  Grenoble Alpes, CNRS, LPMMC, 38000 Grenoble, France}

\author{Aniket Rath}
\affiliation{Univ.  Grenoble Alpes, CNRS, LPMMC, 38000 Grenoble, France}

\author{Petar Jurcevic}
\affiliation{IBM Quantum, Yorktown Heights, NY 10598 USA}

\author{Andreas Elben}
\affiliation{Institute for Quantum Information and Matter, Caltech, Pasadena, CA, USA}
\affiliation{Walter Burke Institute for Theoretical Physics, Caltech, Pasadena, CA, USA}

\author{Cyril Branciard}
\affiliation{Univ. Grenoble Alpes, CNRS, Grenoble INP, Institut N\'eel, 38000 Grenoble, France}

\author{Beno\^it Vermersch}
\affiliation{Univ.  Grenoble Alpes, CNRS, LPMMC, 38000 Grenoble, France}
\affiliation{Institute for Theoretical Physics, University of Innsbruck, Innsbruck, Austria}
\affiliation{Institute for Quantum Optics and Quantum Information of the Austrian Academy of Sciences,  Innsbruck A-6020, Austria}

\date{\today}

\begin{abstract}
We present the experimental measurement, on a quantum processor, of a series of polynomial lower bounds that \textit{converge} to the quantum Fisher information (QFI), a fundamental quantity for certifying multipartite entanglement that is useful for metrological applications.
We combine advanced methods of the randomized measurement toolbox to obtain estimators that are robust against drifting errors caused uniquely during the randomized measurement protocol.
We estimate the QFI for Greenberger–Horne–Zeilinger states, observing genuine multipartite entanglement. Then, we prepare the ground state of the transverse field Ising model at the critical point using a variational circuit.
We estimate its QFI and investigate the interplay between state optimization and noise induced by increasing the circuit depth.
\end{abstract}

\maketitle
The quantum Fisher information is defined with respect to a Hermitian operator $A$ and a quantum state $\rho$, and can be expressed in terms of the spectral decomposition $\rho = \sum_\mu \lambda_\mu \ketbra{\mu}{\mu}$ of the state under consideration as~\cite{Braunstein1994}
\begin{equation}
    F_Q = 2\sum_{(\mu,\nu),\lambda_\mu + \lambda_\nu>0} \frac{(\lambda_\mu - \lambda_\nu)^2}{\lambda_\mu + \lambda_\nu} |\bra{\mu}A\ket{\nu}|^2. \label{eq:qfi}
\end{equation}
It plays a crucial role in various quantum phenomena, including quantum phase transitions~\cite{hauke_measuring_2016,PappalardiQFI2017} and quantum Zeno dynamics~\cite{Smerzi2012} and exhibits profound connections with multipartite entanglement~\cite{Toth2012,Hyllus2012,Pezze2017,Ren2021}.
For $N$ qubits,  with a collective spin operator $A=\frac{1}{2} \sum_{j = 1}^{N} \sigma_{j}^{(\tau)}$~\footnote{$\sigma_{j}^{(\tau)}$ is the Pauli matrix in an arbitrary direction $(\tau)$ acting on the $j^{\rm th}$ spin (identity operators on the other qubits are implicit).}, multipartite entanglement can be certified via QFI in terms of $k$-producibility of the state $\rho$, i.e. a decomposition into a statistical mixture of tensor products of $k$-particle states~\cite{Toth2012,Hyllus2012}. 
The QFI also holds vast applications in resource theory~\cite{Katariya_2021}, many-body physics~\cite{Wang_2014,Macieszczak_2016_dynamical} and quantum metrology~\cite{Pezze2018}.
In particular, for quantum parameter estimation problems, the inverse of the QFI limits the estimation accuracy, as given by the quantum Cramér-Rao bound~\cite{Braunstein1996}. Therefore, it is fundamental for identifying states that provide sensitivities beyond the standard quantum limit~\cite{Pezze2018}.  

In recent years, we have seen an important effort to try to measure the QFI in various experiments (which we will detail in Sec.~\ref{sec:previouswork} below).
This is of interest to test that a quantum device is able to generate nontrivial multipartite entanglement, but also to benchmark the potential of a quantum state for quantum metrology.
In this context, our work provides a measurement of the QFI in a large (up to $13$ qubits) quantum processor.
Note that, while not specifically designed for performing quantum metrology, quantum processors offer, at the moment, unique capabilities for estimating quantum state properties via fast, high-fidelity projective measurements.
The measurement of the QFI is particularly useful for understanding how metrologically relevant quantum states are affected in the presence of unavoidable experimental noise (or what states are more resilient), which in turn can inspire protocols for generating more robust quantum states in actual quantum sensors.  
Importantly, in contrast to previous related approaches accessing lower bounds with finite distance to the QFI, our work provides converging estimations of the QFI. This is particularly relevant for the type of mixed quantum states that are nowadays accessible with current quantum technology, as noisy quantum channels can unpredictably alter the QFI and the previously measured lower bounds. 

Our work is an experimental demonstration of the randomized measurements (RM) protocol presented in  
Ref.~\cite{Rath_QFI_2021}, which proposes a systematic and state-agnostic way to estimate the QFI by measuring a converging series of monotonically increasing lower bounds.
Several practical limitations of this RM protocol have so far prevented the experimental measurement of the QFI:  $(i)$ a prohibitive classical-postprocessing time for reconstructing the QFI from data, $(ii)$ gate and readout errors affecting the RM protocol, and $(iii)$ a prohibitive required number of measurements to overcome statistical errors.
The present work takes advantage of three recently developed methods to adapt Ref.~\cite{Rath_QFI_2021} with respect to the issues $(i)$, $(ii)$, and $(iii)$, and experimentally realize high-fidelity measurements of the QFI. 
First, we use the ``batch shadows'' formalism~\cite{Rath2023OE} to decrease the required post-processing time by several orders of magnitude [issue $(i)$].
Second, we suppress the role of errors occurring in the RM protocol based on the experimental demonstration of robust classical shadows~\cite{Koh2020,Chen2020,Berg2020} [issue $(ii)$]. 
Finally, we apply the formalism of common randomized measurements~\cite{vermersch2023enhanced} to significantly reduce statistical errors compared to the standard RM approach [issue $(iii)$].
This is particularly crucial to obtain a converging value of the QFI with robust classical shadows that typically require more measurements compared to previous approaches~\cite{Koh2020,Chen2020,Berg2020}.
We show, in particular, for the largest system size attainable, that the error mitigation of the RM protocol becomes essential, providing estimates that are compatible with theoretical predictions and up to three times larger than non-mitigated results.
Note that, beyond the measurement of the QFI, we believe that experimentally demonstrating such practical upgrades of the RM toolbox will be useful to measure more faithfully and efficiently other important physical properties for characterizing quantum processors, such as entanglement  entropies~\cite{Brydges2019,Rath2021,satzinger_realizing_2021,Xiao2021,vitale2022symmetry,hoke2023quantum,Joshi2024observing}, negativities~\cite{Elben2020b,Neven2021,carrasco2022entanglement} and state overlaps~\cite{Elben2020a,zhu_cross-platform_2022,joshi2023exploring}.

The manuscript starts with a discussion that puts our results within the framework of previous literature about QFI estimations (Sec.~\ref{sec:previouswork}). 
Our approach is discussed in Sec.~\ref{sec:robust}, where we elaborate on the robust randomized measurement protocol that we implement. 
There, we also discuss the methods introduced above to post-process efficiently the collected experimental RM data~\cite{Rath2023OE,vermersch2023enhanced}.
Finally, in Sec.~\ref{sec:experiment}, we discuss the experimental results where our protocol is applied to measure the converging lower bounds of the QFI for quantum states with up to 13 qubits, prepared on the IBM superconducting device `ibm\_prague'~\cite{IBMQ_ref}.
In particular, we consider two prototypical examples of states: (i) the Greenberger–Horne–Zeilinger (GHZ) state and (ii) the ground state of the transverse field Ising model (TFIM) at the critical point~\cite{QAOAreview}.
Additionally, we provide more details on our work in the Appendices, organized as follows.
In App.~\ref{SM:Fnappendix} and~\ref{sec:depoQFI} we provide the analytical expressions of the lower bounds of the QFI, and analyze their behavior in the presence of global depolarizing errors. In App.~\ref{SM:robust} we give more analytical details of our post-processing protocol and an experimental analysis of the noise mitigation parameters we employ.
In App.~\ref{SM:experimental_checks} we investigate the noise in the platform.
In App.~\ref{SM:locality} we introduce an estimator to verify the locality of the noise and measure it for our experimental setup.
In App.~\ref{moreexpresults} we provide more results for both experiments discussed in the main text.
In App.~\ref{SM:numerics}, we perform a numerical investigation on the statistical error of our estimators to justify the choice of the parameters used in the experiment.

{\renewcommand\addcontentsline[3]{}\section{Previous works on measuring quantum Fisher information and our contribution}\label{sec:previouswork}}

In this section we discuss previous works that estimated the QFI and lower bounds to it. In light of these, we describe how our work goes beyond previous contributions. 
In what follows, we find it useful to distinguish measurement protocols devised in the context of quantum metrology from the tomographic and randomized measurement methods usually associated with quantum processors. The distinction between these two classes of quantum devices is, however, not 100\% sharp, as one can, for instance, consider performing state tomography in a small quantum metrological device based on using local basis transformations.
\vspace{2em}
{\renewcommand\addcontentsline[3]{}\subsection{Quantum metrology methods}}
The QFI was first introduced in the context of quantum metrology for quantifying how accurate the estimation of an unknown parameter $\theta$ could be, and it was readily used to show that the precision of the measurement could go beyond the shot-noise limit (or standard quantum limit)~\cite{Braunstein1996,Pezze2018}.
In quantum metrology, one typically considers realizing the transformation \mbox{$\rho\mapsto\rho(\theta)=e^{-i A \theta}\rho\, e^{iA \theta}$} for some Hermitian operator $A$.
The phase shift $\theta$ is then determined through projective measurements in a given measurement setting, with measurement outcomes $\mu$ coming with probabilities $P(\mu|\theta)$~\cite{Pezze2018}. 
The corresponding uncertainty $\Delta \theta$ is bounded by the classical Fisher information~\cite{fisher1922mathematical,fisher1925theory} (CFI)
\begin{equation}
    F(\theta)=\sum_\mu \frac{1}{P(\mu|\theta)} \left( \frac{\partial P(\mu|\theta)}{\partial \theta}\right)^2,
\end{equation}
according to the Cramér-Rao bound~\cite{Braunstein1994} $\Delta \theta \geq 1/\sqrt{F(\theta)}$. 
The QFI, as defined in Eq.~\eqref{eq:qfi}, is then an upper bound to the CFI and is obtained by maximizing over all possible quantum mechanical measurements~\cite{Braunstein1994}.

In quantum metrological devices, one can thus estimate a lower bound to the QFI by implementing physically the state $\rho(\theta)$, performing projective measurements to estimate $P(\mu|\theta)$ and accessing the CFI $F(\theta)\leq F_Q$ using the above relation~\footnote{Note that $F_Q$ is independent of $\theta$~\cite{Pezze2018}.}. We emphasize here that only when the optimal measurement setting is chosen, the CFI and QFI coincide. 
The CFI has been measured in Refs.~\cite{Strobel2014, Bohnet2016} and has been employed to show genuine multipartite entanglement in GHZ states, for up to $6$ qubits~\cite{Pezze_2016_PNAS}.
In App.~\ref{SM:CFIvsQFI} we provide a numerical comparison of the CFI and QFI for noisy GHZ states. We observe that in the presence of noise, the CFI for some fixed measurement setting provides values that decrease much faster than the QFI as a function of the noise strength, and thus does not represent the optimal metrological content of the prepared state.

Apart from the CFI, various other bounds to the QFI have been proposed and experimentally measured in the quantum metrology context~\cite{Monz2011,Barontini2015,Schmied2016,Pezze2009,Gessner_2019_nonlinear_spinsqueezing,yu2022quantum}. This includes in particular spin-squeezing inequalities that can be directly extracted from measuring expectation values of simple quantum observables. Again, in the presence of unavoidable experimental noise, the distance between a given bound and the QFI is a priori unknown, and it is thus desirable to develop complementary methods to access the QFI directly.

\vspace{2em}
{\renewcommand\addcontentsline[3]{}\subsection{Tomographic and randomized measurements methods}}
Quantum processors such as quantum computers are perhaps not the most natural platform for performing parameter estimation. However, they can be used to experimentally study the generation of quantum states relevant to quantum metrology but also to verify the presence of multipartite entanglement. 
In this context, one can extract an estimate of the QFI via quantum state tomography (QST), and randomized measurement (RM) methods. Note that these methods are also in principle available in quantum metrological platforms, such as cold atoms~\cite{Strobel2014}, but quantum processors, like the superconducting platform employed here, are at the moment more suited for these tasks as they allow for high-fidelity and fast measurement (at the kHz rate). 

QST allows us to reconstruct the state $\rho$ from projective measurements and, therefore, to estimate the QFI by evaluating Eq.~\eqref{eq:qfi}.
For an unknown quantum state, described by density matrices of rank $\chi$, the required number of measurements scales as $\mathcal{O}(\chi^2\, 2^N)$~\cite{O'Donnell2016Jun,Haah2017,flammia2023quantum} and thus is prohibitively demanding for a large number of qubits.
This method has been used to estimate the QFI for small system sizes, up to 4 qubits only~\cite{Krischek_2011_QFI_tomo}.

One can go beyond QST to access QFI in a more scalable way.
Under the assumption of thermal states, one can measure the QFI using dynamical susceptibilities~\cite{hauke_measuring_2016}.
For generic quantum states, one can also rewrite or approximate the QFI in a form that is more suitable for measurements in a quantum processor, i.e., that does not require QST. In recent years, many works proposed nonlinear quantities as a function of the density matrix that lower bounds the QFI~\cite{Rivas2008,Rivas2010,Zhang2017,Girolami2017,Beckey2022,Cerezo2021}.
A notable experimental implementation of this approach was presented in Ref.~\cite{Yu2021} where the authors provide an estimate of a lower bound to the QFI, named sub-QFI~\cite{Cerezo2021}, for GHZ states up to 4 qubits using randomized measurements. As we will explain below, RM protocols require projective measurements in multiple measurements basis, as in QST, but the data is processed to access directly functions of $\rho$, in that case the sub-QFI.
In Ref.~\cite{Yu2021}, RMs were performed on two states $\rho(\theta)$ and $\rho(\theta+d\theta)$, giving access to \mbox{$G(\rho(\theta),\rho(\theta +d\theta))/d\theta^2$} where $G$ is a generalized overlap between quantum states~\cite{Cerezo2021}. 
The protocol suffers from two important limitations: (i) it only estimates the sub-QFI in the limit of small values of $d \theta$ ($d \theta\to 0$), which therefore tends to amplify statistical errors of RMs, (ii) 
the distance between the QFI and the sub-QFI can be significant in the presence of noise.
 
The protocol presented in Ref.~\cite{Rath_QFI_2021}, which we experimentally implement here, addressed these two challenges. To be specific, we use RMs to access a converging series of lower bounds of the QFI, i.e., not only one bound to the QFI. In addition, our estimators do not rely on measuring asymptotically a relative overlaps between two states  $\rho(\theta)$ and $\rho(\theta+d\theta)$. This allows us to reach a number of qubits of $13$, i.e., more than three times the system size that was achieved so far with QST~\cite{Krischek_2011_QFI_tomo}.
In App.~\ref{SM:DGvsQFI}, we compare our experimental estimations of the QFI with the estimator in Ref.~\cite{Yu2021}, which we show we can extract faithfully from a single RM experiment.

\begin{figure*}
    \centering
    \includegraphics[width =\linewidth]{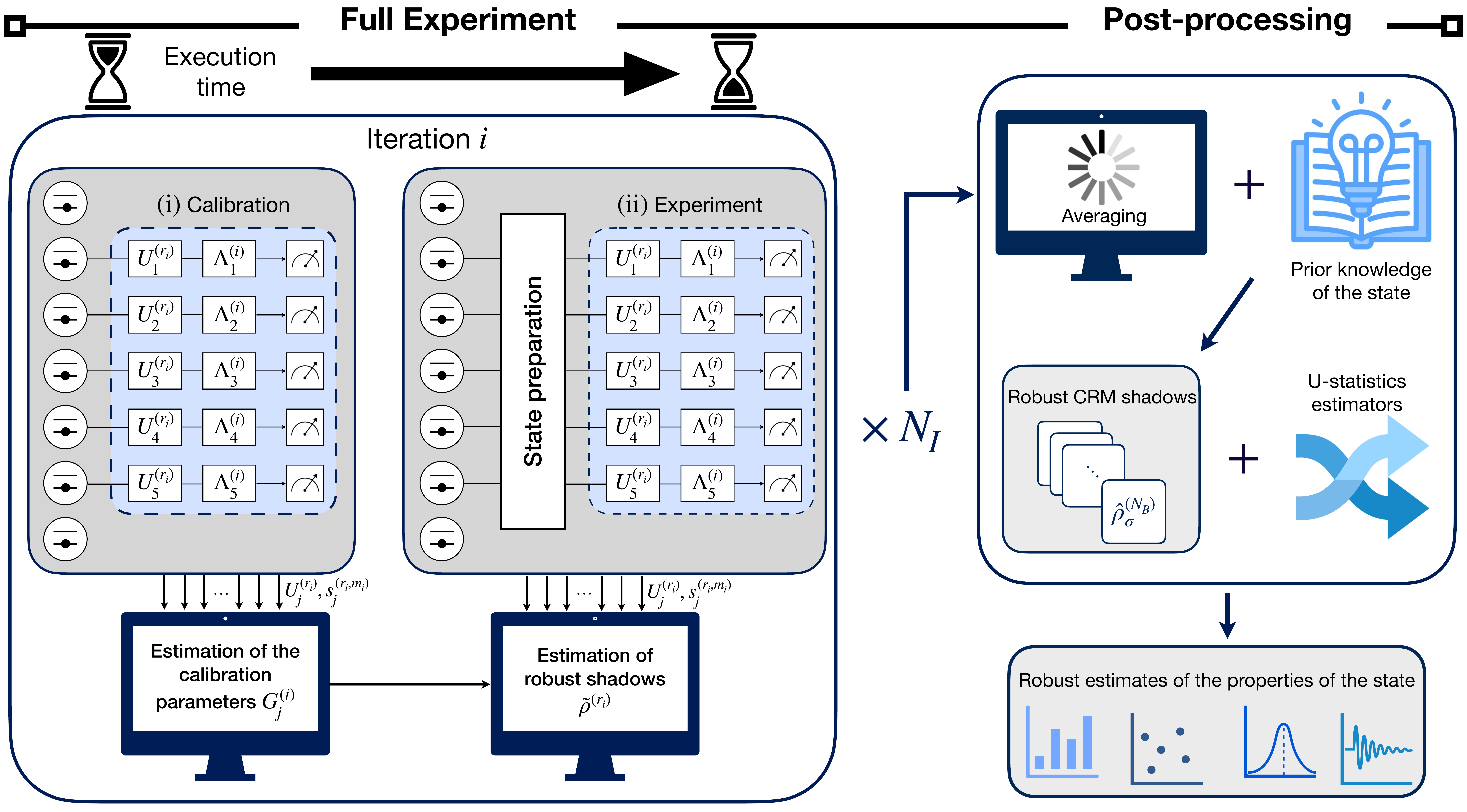}
    \caption{Overview of experimental protocol and post-processing. The experimental protocol is comprised of several `iterations' $i=1,\dots,N_I$. Each of them consists of a calibration step (i) and the experiment on the state of interest $\rho$ (ii). From the unitaries and bit-strings recorded in (i) we estimate the noise parameter $G_j^{(i)}$ (Eq.~\eqref{eq:GMT}), that is used for the construction of the robust shadows $\tilde{\rho}^{(r_i)}$ (Eq.~\eqref{eq:robshadows}) together with the data from step (ii). Integrating an approximation $\sigma$ of the state $\rho$, all robust shadows are then collected and averaged in batches to obtain $b=1,\dots,N_B$ robust common randomized batch shadows $\hat{\rho}^{(b)}_{\sigma}$ (Eq.~\eqref{eq:batchshadows}). These are used for computing the multi-copy observables of interest through the U-statistics estimator (Eq.~\eqref{eq:batchestimators}). Our experiments are performed on the IBM superconducting qubit device `ibm\_prague'~\cite{IBMQ_ref}.}
    \label{fig:fig1} 
\end{figure*}

\vspace{2em}
{\renewcommand\addcontentsline[3]{}\section{Estimation of multi-copy observables from noisy randomized measurement data} \label{sec:robust}}
In this section we describe all the building blocks for the estimation of multi-copy observables employing RM data. In particular, we present robust estimators of the QFI, the purity and the fidelity of a quantum state, and detail the needed post-processing protocol to mitigate the noise and reduce statistical errors.

\vspace{2em}
{\renewcommand\addcontentsline[3]{}\subsection{Data acquisition with randomized measurements}\label{sec:EP}}

Our approach, illustrated in Fig.~\ref{fig:fig1}, is comprised of several repetitions of two building blocks: (i) Calibration of randomized measurements and (ii) randomized measurements on the state of interest $\rho$.
The calibration step 
is employed to learn and mitigate the gate and readout errors that affect the measurements, as described in Refs.~\cite{Chen2020, Koh2020,Berg2020}.
This relies on the ability to prepare on the experimental platform a specific state with high fidelity.  In this work, we fix the calibration state to be $\ket{\mathbf{0}} \equiv \ket{0}^{\otimes N}$, which is producible with high efficiency on our quantum processor. 
The data collected in step (ii) are then used for estimating the observables we are interested in. 
We call each run of (i) and (ii) an `iteration' of the experiment. Performing consecutive iterations allows us to account for the temporal variations in gate and readout errors. 
Assuming that the temporal fluctuations of the errors affecting the randomized measurements protocol for each iteration are negligible, each calibration step captures the specific error profile of a distinct time window within the overall experimental run. 

Let us start by recalling the randomized measurement protocol in the absence of noise.
We begin by preparing the $N$-qubit quantum state $\rho$. Then we apply local random unitaries $U = U_1 \otimes \dots \otimes U_N$ where the local (single-qubit) unitaries $U_j$ ($j=1,\dots,N$) are sampled from the circular unitary ensemble~\cite{mezzadri2007generate}. The rotated state $U \rho U^\dag$ is then projected onto a computational basis state $\ket{\s} = \ket{s_1, \dots, s_N}$, where $s_j \in \{0,1\}$ for $j=1,\dots,N$, by performing a measurement.
To make the protocol robust against the noise occurring in the quantum device, we apply the measurement sequence described above on the states $\ket{\mathbf{0}}$, $\rho$ in steps (i), (ii) of Fig.~\ref{fig:fig1}, respectively.
As mentioned before, the data collected from (i) is used to mitigate the errors induced by the noisy measurement protocol in step (ii)~\cite{Chen2020,Koh2020}.

\vspace{2em}
{\renewcommand\addcontentsline[3]{}\subsection{Assumptions on the noise affecting the randomized measuements}}
The basic assumptions on the noise model for our post-processing protocol are as follows.
As in Ref.~\cite{Chen2020}, we consider a gate-independent noise channel $\Lambda$, applied after the random unitaries: that is, for each chosen $U$ the state $\rho$ transforms as $\Lambda(U \rho U^{\dagger})$.
We assume that the noise channel $\Lambda$ is constant during each iteration $i$ -- we label it as $\Lambda^{(i)}$ -- and may change between each iteration. 
We provide experimental evidence of the variation of the noise over different iterations -- that is remarkably captured by our protocol -- in App.~\ref{SM:batchexp_check} .
Additionally, we assume the noise  to be local for each qubit, so that $\Lambda^{(i)}=\Lambda^{(i)}_1\otimes \dots \otimes \Lambda^{(i)}_N$.
In App.~\ref{SM:locality} we provide and implement a method to verify the assumption of locality of the noise, based on the calibration data. 
Additionally, in App.~\ref{SM:batchvsfull}, we show that tracking the variation of the noise over the different iterations is essential to provide faithful estimations of the QFI.

\vspace{2em}
{\renewcommand\addcontentsline[3]{}\subsection{Robust classical shadows}}
The first step towards the measurement of non-local observables of interest is to construct estimators of the density matrix $\rho$ from the noisy measurements.
This object, called a `robust shadow'~\cite{Chen2020} (see also App.~\ref{SM:rob_from_meas}), can be defined as
\begin{align}
& \tilde{\rho}^{(r_i)} \notag \\
& \!= \sum_{\s} \hat{P}(\s|U^{(r_i)})\bigotimes_{j=1}^N \left( \alpha^{(i)}_j\;{U^{(r_i)}_j}^{\dag} \ketbra{s_j}{s_j} U^{(r)}_j- \beta^{(i)}_j\mathbb{1}\right), \label{eq:robshadows}
\end{align}
where $\alpha^{(i)}_j= \frac{3}{2 G^{(i)}_j-1}$ and $\beta^{(i)}_j = \frac{G^{(i)}_j-2}{2 G^{(i)}_j-1}$.
Here $r_i$ labels a unitary in iteration $i$ and $\hat{P}(\s|U^{(r_i)}) = \sum_{m_i = 1}^{N_M} \frac{\delta_{\s, \s^{(r_i,m_i)}}}{N_M}$ is the estimated (noisy) Born probability, where $m_i$ labels an individual measurement performed after the application of $U_j^{(r_i)}$, whose outcome is the bit-string denoted $\s^{(r_i,m_i)}$. 
The quantity in Eq.~\eqref{eq:robshadows} satisfies $\mathbb{E}[\tilde{\rho}^{(r_i)}] = \rho$, where the average is taken over the applied unitaries. 
This equality is necessary in particular to derive the unbiased estimators of the lower bounds of the QFI~\cite{Rath_QFI_2021}.

The quantity $G^{(i)}_j$ introduced above contains the relevant information about the noise on qubit $j$  in iteration $i$ of the measurement protocol. It is defined as
\begin{equation}\label{eq:GMT}
G^{(i)}_j
=
\frac{1}{2}\sum_{s_j=0,1}\bra{s_j} \Lambda^{(i)}_{j}(\ketbra{s_j}{s_j}) \ket{s_j}
\end{equation}
and can be interpreted as the average `survival probability' of the two basis states of qubit $j$. In the absence of noise, $G^{(i)}_j=1$  ($\forall\; j=1,\dots,N$), and one recovers the standard `classical shadow'~\cite{Huang2020}. In the opposite limit of fully depolarising noise, $G^{(i)}_j=1/2$, the coefficient $\alpha^{(i)}_j,\beta^{(i)}_j$ diverge and the estimators suffer from large statistical errors~\cite{Chen2020}. In our work $G^{(i)}_j \sim 0.98$ (See App.~\ref{SM:comparisonG}).

For each iteration $i$ and each qubit $j$, $G^{(i)}_j$ is computed through the experimental data collected during the calibration step.
As detailed in App.~\ref{SM:calibration}, we can define the unbiased estimator
\begin{equation}\label{eq:estimatorCjSM}
\hat{G}^{(i)}_j =\frac{3}{N_U}\sum_{r_i,s_j} \hat{P}(s_j|U_j^{(r_i)}) P(s_j|U_j^{(r_i)})-1.
\end{equation}
Here, $\hat{P}(s_j|U_j^{(r_i)}) = \sum_{m_i = 1}^{N_M} \frac{\delta_{s_j, s_j^{(r_i,m_i)}}}{N_M}$ is the estimated (noisy) Born probability and $P(s_j|U_j^{(r_i)}) = |\bra{s_j} U^{(r_i)}_j\ket{0}|^2$ is the theoretical (noiseless) one for a single qubit $j$. The information on the noise is contained in $\hat{P}(s_j|U_j^{(r_i)})$, that approaches the theoretical noisy Born probability $P_\Lambda(s_j|U_j^{(r_i)})=\bra{\s} \Lambda(U^{(r_i)}\rho\, {U^{(r_i)}}^\dag ) \ket{\s}$ in the limit $N_M\rightarrow \infty$.
We remark here that all our results are compatible with the ones in Ref.~\cite{Chen2020}, where a slightly different formalism has been employed.

\vspace{2em}
{\renewcommand\addcontentsline[3]{}\subsection{Common randomized shadows and noise estimator}\label{subsec:crm}}
The statistical errors can be significantly reduced by using \emph{common randomized measurements}~\cite{vermersch2023enhanced} to define an estimator with smaller variance.
The central idea is to construct the robust shadows integrating an approximation of the state $\rho$ in the form of some classical representation $\sigma$. In practice, we consider $\sigma$ to be the ideal pure state that we intend to realise in our experiment. We build `common randomized' (CRM) shadows as
\begin{equation}
    \tilde{\rho}_\sigma^{(r_i)}
    =
    \tilde{\rho}^{(r_i)}
    -\sigma^{(r_i)}
    +\sigma ,
    \label{eq:shiftedshadow}
\end{equation}
where the term $\sigma^{(r_i)}$ is constructed from $\sigma$ as  
 \begin{equation}
    \sigma^{(r)}
    =\sum_{\s} P_{\sigma}(\s|U^{(r_i)})\bigotimes_{j=1}^N \left( 3\;{U^{(r)}_j}^{\dag} \ketbra{s_j}{s_j} U^{(r_i)}_j- \mathbb{1}\right) \label{eq:def_sigma_r}
 \end{equation}
 with
 ${P}_\sigma(\s|U^{(r_i)})=\bra{\s}U^{(r_i)} \sigma {U^{(r_i)}}^\dag\ket{\s}$. The latter is a fictitious probability distribution obtained from computational basis measurements on the $\sigma$ rotated by the same unitaries applied in the experiment $U^{(r_i)}$, done on a classical device. One may notice that $\expp[ {\sigma}^{(r_i)} ] = \sigma$~\cite{vermersch2023enhanced}.
 Thus, $\hat{\rho}_\sigma^{(r_i)}$ is an unbiased estimator of $\rho$, as 
 \begin{equation}
     \expp[ \hat{\rho}_\sigma^{(r_i)}]=\rho -\sigma + \sigma  = \rho, 
 \end{equation} 
 irrespective of the choice of $\sigma$.
 Crucially, this procedure enters entirely during post-processing thus leaving the data acquisition of the experiment independent from it.

The same reasoning of common randomized numbers~\cite{vermersch2023enhanced}
can be used to improve the statistical accuracy of  the estimator of $G^{(i)}_j$. 
Let us introduce for that the quantity
\begin{equation}\label{eq:Bmaintext}
B^{(i)}_j =\frac{3}{N_U}\sum_{r_i,s_j}P(s_j|U_j^{(r_i)})^2.
\end{equation}
With this, we then define
\begin{equation}\label{eq:CCRestimator}
\hat{\mathcal{G}}^{(i)}_j=\hat{G}^{(i)}_j-B^{(i)}_j + \mathbb{E}[B^{(i)}_j]. 
\end{equation}
Here, $\hat{\mathcal{G}}^{(i)}_j$ and $\hat{G}^{(i)}_j$ have the same expectation value, but the variance of $\hat{\mathcal{G}}^{(i)}_j$ is smaller because $\hat{G}^{(i)}_j$ and $B^{(i)}_j$ are positively correlated.
Observing in particular that $\mathbb{E}[B^{(i)}_j] = 2$, see App.~\ref{SM:calibration}, we can then write the new estimator as
\begin{align}\label{eq:estimatorMT}
\hat{\mathcal{G}}^{(i)}_j
=
\frac{3}{N_U}
\sum_{r_i,s_j}
\widehat{\Delta P}(s_j|U_j^{(r_i)})
P(s_j|U^{(r_i)}_j)
+1,
\end{align}
where $\widehat{\Delta P}(s_j|U_j^{(r_i)}) = \hat P(s_j|U^{(r_i)}_j) - P(s_j|U^{(r_i)}_j)$ is the difference between the experimentally estimated (noisy) Born probability and the theoretical (noiseless) one.
The fact that $\hat{\mathcal{G}}^{(i)}_j$ is a more efficient estimator of the noise term $G^{(i)}_j$ is shown explicitly in App.~\ref{SM:comparisonG}, employing our experimental data.

\vspace{2em}
{\renewcommand\addcontentsline[3]{}\subsection{Data compression and U-stastics estimators}\label{sec:datacompression}}
The last step of our protocol consists in compressing the CRM shadows in order to minimize the post-processing time. To do so, we compress the $N_I$ estimators $\tilde{\rho}^{(r_i)}$ into $N_B$ `robust CRM batch shadows'~\cite{Rath2023OE} $\hat{\rho}_{\sigma}^{(b)}$ as (we assume $N_B$ divides $N_I$):
\begin{equation}\label{eq:batchshadows}
    \hat{\rho}^{(b)}_{\sigma}= \frac{N_B}{N_I}\sum_{i=(b-1)N_I/N_B+1}^{bN_I/N_B} \left(\sum_{r_i} \frac{\tilde{\rho}_{\sigma}^{(r_i)}}{N_U}\right)
\end{equation}
for $b=1,\ldots,N_B$. 
The latter can be used to estimate any multi-copy observable of interest, i.e. functions $f_n$ in the form \mbox{$f_n = \tr(O^{(n)} \rho^{\otimes n})$}. Given the $N_B$ robust CRM batch shadows $\hat{\rho}_{\sigma}^{(b)}$, one can provide an unbiased estimator of the function $\hat{f}_n$ using U-statistics~\cite{Hoeffding1992}. 
This is achieved by replacing each copy of the density matrix in the multi-copy function $f_n$ with a different robust CRM batch shadow and computing the average over all possible such choices. 
This is explicitly expressed as
\begin{equation}
    \hat{f}_n = \frac{1}{n!\binom{N_B}{n}} \sum_{b_1 \ne \dots \ne b_n} \Tr\left ( O^{(n)} \hat{\rho}_{\sigma}^{(b_1)}  \otimes \dots \otimes \hat{\rho}_{\sigma}^{(b_n)} \right), \label{eq:Ustat}
\end{equation}
where $\rho^{\otimes n}$ from $f_n = \Tr(O^{(n)} \rho^{\otimes n})$ is replaced by an average over $\hat{\rho}_{\sigma}^{(b_1)} \otimes \dots \otimes \hat{\rho}_{\sigma}^{(b_n)}$ with $b_1 \ne \dots \ne b_n$. Such estimators can be evaluated with a classical post-processing which scales with the number of elements to be evaluated in the sum. Then, it is clear that the compression of the data from $N_I$ to $N_B$ shadows allows for faster post-processing time, that changes from $\mathcal{O}(N_I^n)$ to $\mathcal{O}(N_B^n)$. This comes at the expense of storing large dense matrices, albeit not compromising the statistical performances~\cite{Rath2023OE}. Finally, as thoroughly explained in Ref.~\cite{Rath2023OE}, one has to consider that while the statistical accuracy increases with $N_B$, so does also the post-processing time. Then one has to find an $N_B$ that provides a good balance between good statistical performances and reasonable post-processing cost.

\vspace{2em}
{\renewcommand\addcontentsline[3]{}\subsection{Estimators of the QFI as a converging series of polynomials}}
Let us now discuss in detail the estimators of the QFI $F_Q$ that we use.
As shown in Ref.~\cite{Rath_QFI_2021}, while the QFI cannot be accessed directly by randomized measurements, as written in Eq.~\eqref{eq:qfi}, it can be alternatively expressed and estimated in terms of a converging series of monotonically increasing lower bounds $F_n$.
For the first three orders $n = 0,\, 1,\, 2$, one can write explicitly
\begin{align}\label{eq:Fseries}
    &&
    F_0 &= 4\,\mathrm{Tr}(\rho \big[\rho,A\big] A) \,,  \nonumber \\
    &&
    F_1 &= 2\,F_0 - 4\,\Tr(\rho^2[\rho,A]A) \,, \\
    &&
    F_2 &= 3(F_1 - F_0) +4\,\Tr(\rho^2 [\rho^2,A]A) \,, \nonumber
\end{align} 
where $[\cdot,\cdot]$ is the commutator.
We provide the general expression for $F_n$ in App.~\ref{SM:Fnappendix}.
Each function $F_n$ is a polynomial function of the density matrix (of order $n+2$); such functions can now be accessed via randomized measurements, as it has been shown for  entropies~\cite{Brydges2019,Rath2021,satzinger_realizing_2021,Xiao2021,vitale2022symmetry,hoke2023quantum},  negativities~\cite{Elben2020b,Neven2021,carrasco2022entanglement}, state overlaps~\cite{Elben2020a,zhu_cross-platform_2022,joshi2023exploring}, scrambling~\cite{Manojo2020} and topological invariants~\cite{Cian2021,Elben2019}.

We define unbiased estimators $\hat{F}_n$ for the lower bounds ${F}_n$ according to the rules of U-statistics~\cite{Hoeffding1992,Huang2020} by summing over all possible disjoint indices, as in Eq.~\eqref{eq:Ustat} above. 
In practice~\cite{Rath_QFI_2021}, for $n = 0,\, 1,\, 2$ one can thus write (assuming $N_B>n+2$)
\begin{equation}\label{eq:batchestimators}
\begin{aligned}
    \hat{F}_0 &= {\textstyle \frac{4(N_B-2)!}{N_B!}} \sum_{b_1 \ne b_2} \Tr \left( \hat{\rho}_{\sigma}^{(b_1)} [\hat{\rho}_{\sigma}^{(b_2)}, A] A \right), \\
    \hat{F}_1 &= 2 \hat{F}_0 - {\textstyle \frac{4(N_B-3)!}{N_B!}} \sum_{b_1\ne \dots \ne b_3} \Tr \left( \hat{\rho}_{\sigma}^{(b_1)}\hat{\rho}_{\sigma}^{(b_2)} [\hat{\rho}_{\sigma}^{(b_3)}, A] A \right), \\
    \hat{F}_2 &= 3( \hat{F}_0 {-}  \hat{F}_1) {+} {\textstyle \frac{4(N_B-4)!}{N_B!}} \!\!\!\!\! \sum_{b_1\ne \dots \ne b_4}\!\!\!\!\!\! \Tr \!\left( \!\hat{\rho}_{\sigma}^{(b_1)} \!\hat{\rho}_{\sigma}^{(b_2)}[\hat{\rho}_{\sigma}^{(b_3)}\!\hat{\rho}_{\sigma}^{(b_4)}\!, A] A \!\right)\!.
\end{aligned}
\end{equation}
where $[\cdot, \cdot]$ denotes the commutator and we choose $N_B=10$.
The estimators $\hat{F}_n$ suffer from statistical errors arising due to the finite number of unitaries and measurements performed. Even though $\hat{F}_n$ exponentially converges to the true value of QFI as a function of the order $n$ of the bound, the statistical error on the estimator increases with $n$ for a fixed measurement budget~\cite{Rath_QFI_2021}. In App.~\ref{SM:measbudget} we show the scaling of the required number of measurements for a given value of statistical error $\mathcal{E}$ as a function of the system size $N$. 
Accurate variance bounds for $\hat{F}_n$ have been discussed in Ref.~\cite{Rath_QFI_2021}.

\vspace{2em}
{\renewcommand\addcontentsline[3]{}\subsection{Estimators of fidelity and purity}}
As anticipated, the protocol presented in this manuscript is not restricted to the estimation of the lower bounds $F_{n}$. For example, we can extract two other important quantities: fidelity with respect to an ideal state, and purity.
We write here the estimators in terms of the CRM batch shadows $\hat{\rho}_{\sigma}^{(b)}$.
Following Eq.~\eqref{eq:Ustat} once again, the estimator of the purity can be expressed as
\begin{equation}
    \widehat{\Tr\left(\rho^2\right)}= \frac{1}{N_B(N_B-1)} \sum_{b_1 \ne b_2} \Tr \left( \hat{\rho}_{\sigma}^{(b_1)} \hat{\rho}_{\sigma}^{(b_2)}\right).
\end{equation}
On the other hand, assuming that an ideal state is described by the density matrix  $\sigma'$, one can estimate the overlap of the latter with the prepared state $\rho$ as
\begin{equation}
    \widehat{\Tr\left( \rho \sigma' \right)}=\frac{1}{N_B} \sum_{b} \Tr \left( \hat{\rho}_{\sigma}^{(b)} \sigma'\right).
\end{equation}
In the following, since we are interested in the quality of the state preparation on the hardware, we will testify it by measuring these two quantities.

\vspace{2em}
{\renewcommand\addcontentsline[3]{}\section{Experimental results} \label{sec:experiment}}
In this section we describe the experimental results that were obtained on IBM superconducting processors. As mentioned before, we will consider two states: the GHZ state in Sec.~\ref{sec:GHZstate} and the ground state of the transverse field Ising model (TFIM) at the critical point in Sec.~\ref{sec:QAOA}. In our work, the observable under consideration is taken to be $A=\frac{1}{2}\sum_j \sigma_j^z$, where $\sigma_j^z$ is the Pauli-$z$ operator acting on qubit $j$. 

\vspace{2em}
{\renewcommand\addcontentsline[3]{}\subsection{Measurement budget}\label{sec:budget}}
The full experiment is divided into a total of $N_I$ iterations (labeled by $i=1,\dots,N_I$). For steps (i) and (ii) in each iteration $i$, we apply the same \mbox{$N_U = 200$} local random unitaries \mbox{$U^{(r_i)} = U_1^{(r_i)} \otimes \dots \otimes U_N^{(r_i)}$}, with \mbox{$r_i= 1, \dots, N_U$}~\footnote{One can also implement different unitaries for calibration and estimation. For the present experiment, we did not notice significant differences as we estimated different quantities.}, and (for each unitary) record $N_M = 1000$ measurement outcome bit-strings \mbox{$\s^{(r_i,m_i)} = \left(s^{(r_i,m_i)}_1,\dots , s^{(r_i,m_i)}_N\right)$} with \mbox{$m_i = 1, \dots, N_M$}. 

The total measurement budget ($N_I N_U N_M$) that is required to reach a given accuracy for an estimator depends on the size of the system $N$~\cite{elben_randomized_2023}.
In particular, for our experiments, we employ a total of  \mbox{$N_U^{\textrm{tot}} = N_I N_U = 300 \cdot 2^{0.5 N}$} unitaries to obtain an estimation error of \mbox{$\sim 10\%$} on the highest-order estimated lower bound of the QFI (without exploiting any prior knowledge of the prepared quantum state).
Note that the higher the order, the more measurements are needed to overcome statistical fluctuations. Numerical investigations on the measurement budget are detailed in App.~\ref{SM:measbudget}.

\vspace{2em}
{\renewcommand\addcontentsline[3]{}\subsection{GHZ states}\label{sec:GHZstate}}

\begin{figure*}
    \centering\includegraphics[width=0.9\linewidth]{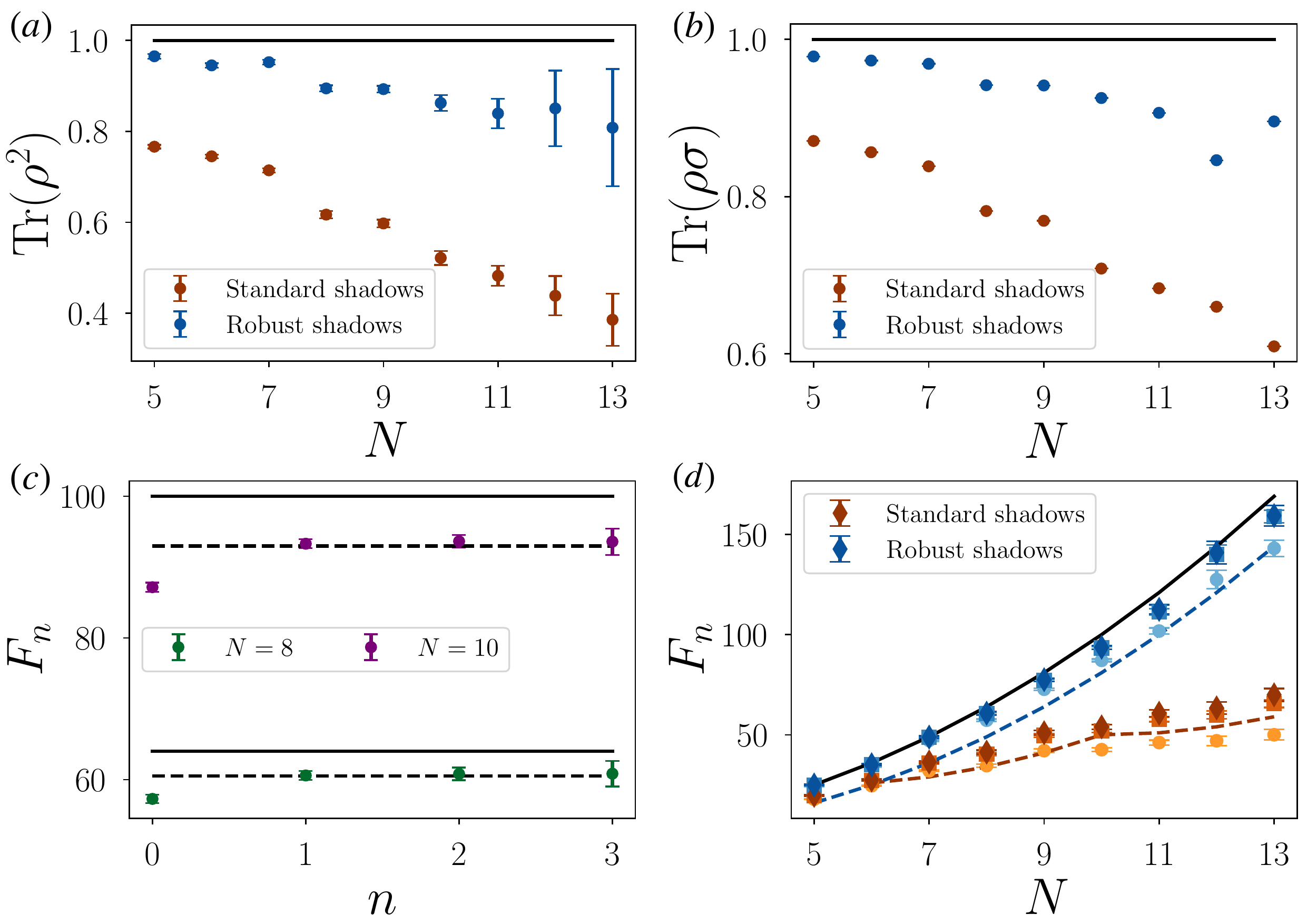}
    \caption{Experimental results for the pure GHZ state. $(a)$ Purity of the prepared state. $(b)$ Fidelity with respect to the GHZ state $\sigma=\ketbra{\rm GHZ}{\rm GHZ}$. $(c)$ Convergence of  $F_n$ as a function of $n$ for $N=8,10$ qubits (green and violet respectively) to the true value of QFI. Solid lines represent the theoretical value for pure GHZ states, dashed lines denote the case of noise affecting the system in the form of global depolarization. $(d)$ $F_0$, $F_1$, $F_2$ (light to dark with circle, square, and diamond respectively) as a function of the number of qubits $N$. The solid line is the exact value of the QFI $F_Q = N^2$ for pure GHZ states. The dashed blue line corresponds to the entanglement witness \mbox{$\Gamma (N,k=N-1) = (N-1)^2$} above which the state is genuinely multipartite entangled.
    The dashed orange line corresponds to the entanglement witness $\Gamma (N,k=5)$ above which we detect a state to be at least 6-partite entangled.
    The details of the experimental protocol are described in Sec. \ref{sec:budget}.}
    \label{fig:fig2}
\end{figure*}
The Greenberger-Horne-Zeilinger (GHZ) state is a fundamental resource for various quantum information processing tasks, including quantum teleportation~\cite{Boschi1998,bouwmeester1997experimental}, quantum error correction~\cite{gottesman1999demonstrating,knill1998resilient}, and quantum cryptography~\cite{Popescu1992}.
It can be written as
\begin{equation}
    \ket{\text{GHZ}} = \frac{1}{\sqrt{2}}\left(\ket{0}^{\otimes N} + \ket{1}^{\otimes N} \right).
\end{equation}
Remarkably, GHZ states are ideal candidates for quantum metrology as they saturate the value of the QFI \mbox{$(F_Q = N^2)$} and, thus, can be used to reach higher sensitivities in parameter estimation that scale as $\sim N^{-1}$ (known as the \emph{Heisenberg limit}), and is beyond the standard shot-noise limit $\sim N^{-1/2}$~\cite{Giovannetti2006,Toth2012,Toth2014}.

By implementing randomized measurements, we experimentally estimate the QFI as a function of different system sizes $N$ and witness the presence of multipartite entanglement~\cite{Toth2012,Hyllus2012,Ren2021}. 
An $N$-qubit pure GHZ state is genuinely multipartite entangled (GME), i.e., it cannot be decomposed into a statistical mixture of tensor products of $(N-1)$-particle states.
In general, one can use the inequality $F_Q > \Gamma(N,k)$, with \mbox{$\Gamma(N,k) = \floor*{\frac{N}{k}}k^2 + \big(N - \floor*{\frac{N}{k}}k\big)^2$}, to certify  that a state is not $k$-producible, i.e., that it has an `entanglement depth' of at least $k+1$~\cite{Toth2012,Hyllus2012}. In this case, it is said to be $(k+1)$-partite entangled. The inequality is particularly relevant in the presence of noise, where a perfect pure state is not produced.
Until now, fidelity measurements have allowed validating GME in GHZ states prepared on superconducting qubits~\cite{Song_10_GHZ_2017, Mooney_2021}, 14 trapped ions~\cite{Monz2011}, 18 photonic qubits~\cite{wang_18_GHZ_2018} and other multipartite entangled states~\cite{Song_2019,Gong_2019,Pogorelov_2021,cao_generation_2023}.
Additionally, GME states can also be verified via multiple coherences for GHZ states~\cite{omran2019generation,Wei2020}. 

We show our experimental results in Fig.~\ref{fig:fig2}.
As we mentioned earlier, with the RM framework,
we can access many interesting quantum
properties from the same experimental dataset. First, to test the quality of the state preparation on our quantum hardware, we extract two important quantities, namely the purity of the final state ($\Tr(\rho^2)$) and the fidelity ($\Tr(\rho \sigma)$) with respect to a pure GHZ state $\sigma = \ketbra{\rm GHZ}{\rm GHZ}$. 
We plot these results in Fig.~\ref{fig:fig2}$(a)$ and in Fig.~\ref{fig:fig2}$(b)$ respectively. 
For each panel, the blue points denote the experimental results in the case error mitigation is performed, while the orange ones correspond to the case when it is not, i.e. we take $G^{(i)}_j=1$, $\alpha^{(i)}_j=3$ and $\beta^{(i)}_j=-1$ in Eq.~\eqref{eq:robshadows}.
We observe clearly that the robust protocol mitigates the errors occurring during the measurement phase as indicated by higher values of fidelities (between $0.85 - 0.97$) and purities (between $0.8 - 0.95$) compared to the unmitigated results for the whole qubit range.
In both cases, we observe the decreasing trend of the fidelity and the purity as a function of the system-size $N$. 
This signature clearly indicates that noise is induced during state preparation of the GHZ states as the two-qubit gate count increases with $N$.

Let us analyze now the convergence of the lower bounds to the QFI, in Fig.~\ref{fig:fig2}$(c)$.
Here we plot $F_n$ ($n=0,1,2,3$) for the GHZ state prepared on a system of $N=8$ (green) and $N = 10$ (violet) qubits. In the absence of noise, the theoretical value of the QFI for a GHZ state is $F_Q=N^2$.
This is plotted as a thick black line for both  $N=8$ and $N = 10$. From the experimental results, it is clear that the lower bounds do not converge to $N^2$, but to a lower value.
We can understand it considering that the state preparation is affected by noise, as it is suggested by the measurements of purity and overlap.
One simple way for modeling this is by assuming that the system is affected by global depolarization, turning the pure GHZ state into $\rho = (1-p_D) \ketbra{\rm GHZ}{\rm GHZ} + p_D \, \mathbb{1}/2^N$. The noise parameter $p_D$ can be extracted from the experimental values of the purity according to the following relation 
\begin{equation}
    \Tr(\rho^2) = (1-p_D)^2 + \frac{2p_D - p_D^2}{2^N}.
\end{equation}
With this specific noise model, one can better understand the convergence of the lower bounds $F_n$ to a finite value of QFI as a function of the bound order $n$. 
While in the noiseless scenario one has $F_Q = N^2$, the theoretical value of the QFI of the noisy GHZ state is found to be~\cite{Hyllus2012} 
\begin{equation}\label{eq:qfidepoMT}
F_Q(p_D) = N^2 \frac{(1-p_D)^2}{1-p_D+2p_D/2^N}
\end{equation}
For $N=8$ and $N=10$ qubits, they correspond to the values of $F_Q(p_D) = 60.3 \pm 0.45$ and $F_Q(p_D) = 92.7 \pm 1.83$. The respective values of $p_D$ are $0.056 \pm 0.007$ and $0.072 \pm 0.018$, which are extracted from the mitigated values of purities, i.e. $0.89 \pm 0.0066$ and $0.86 \pm 0.017$ (See App.~\ref{sec:depoQFI}). 
In Fig.~\ref{fig:fig2}$(c)$ we draw $F_Q$ and $F_Q(p_D)$ as solid and dashed black lines, respectively. 
We observe the convergence of $F_n$ to the values of the QFI for the extracted value of $p_D$ within error bars of the estimations, for both values of $N$. As mentioned in Sec.~\ref{sec:datacompression}, we observe that, even though $F_n$ exponentially converges to the QFI as
a function of $n$, its statistical error increases at a fixed measurement budget. This is thoroughly discussed in Ref.~\cite{Rath2023OE}.

In Fig.~\ref{fig:fig2}$(d)$ we show the experimental measurements of $F_0, \, F_1 $, and $F_2$ (light to dark) on the prepared GHZ state as a function of $N$.
The black thick line provides the ideal scaling of the QFI ($F_Q = N^2$) for pure GHZ states. The black dashed line, instead, denotes the entanglement witness $\Gamma(N,k=N-1)$ that scales as $(N-1)^2$ and above which we can consider our prepared states to be GME.  
The experimental points correspond to the measured values for two different cases: mitigated results through our calibration step in blue, and raw data without performing the calibration step in orange. We observe that the mitigated data used to estimate $F_n$ violates the necessary entanglement witness to be GME for any size $N$, hence all our prepared states have an entanglement depth of $k = N$.
Thus, we demonstrate the presence of multipartite entanglement through the estimation of \emph{converging} lower bounds to the QFI, whose convergence to the true value has been shown in Fig.~\ref{fig:fig2}$(a)$. 

Analyzing the raw data (orange points in Fig.~\ref{fig:fig2}$(b)$) that are prone to errors during the randomized measurement protocol gives us lower estimations of the bounds.
They do not violate the GME threshold and do not follow the expected scaling seen for the mitigated data points.
This shows that the error mitigation in the measurement protocol is decisive and useful for estimating the underlying properties of the prepared quantum states.
In the case of the analysis of the raw data, we can assert from the witness bound $F_Q > \Gamma(N,k)$~\cite{Toth2012, Hyllus2012} that our prepared state contains an entanglement depth of $k = 6$ for $N\ge 6$. 
Importantly, in App.~\ref{SM:batchvsfull}, we show the estimation of the lower bounds in the case when the calibration (step (i) in Fig.~\ref{fig:fig1}) is done entirely at the beginning and is followed by the experiment (step (ii) in Fig.~\ref{fig:fig1}). We observe clearly that performing the calibration in multiple iterations provides better results (closer to the theoretical values) for larger system sizes where the full experimental duration starts to increase.

\begin{figure}
    \centering
    \includegraphics[width=0.9\linewidth]{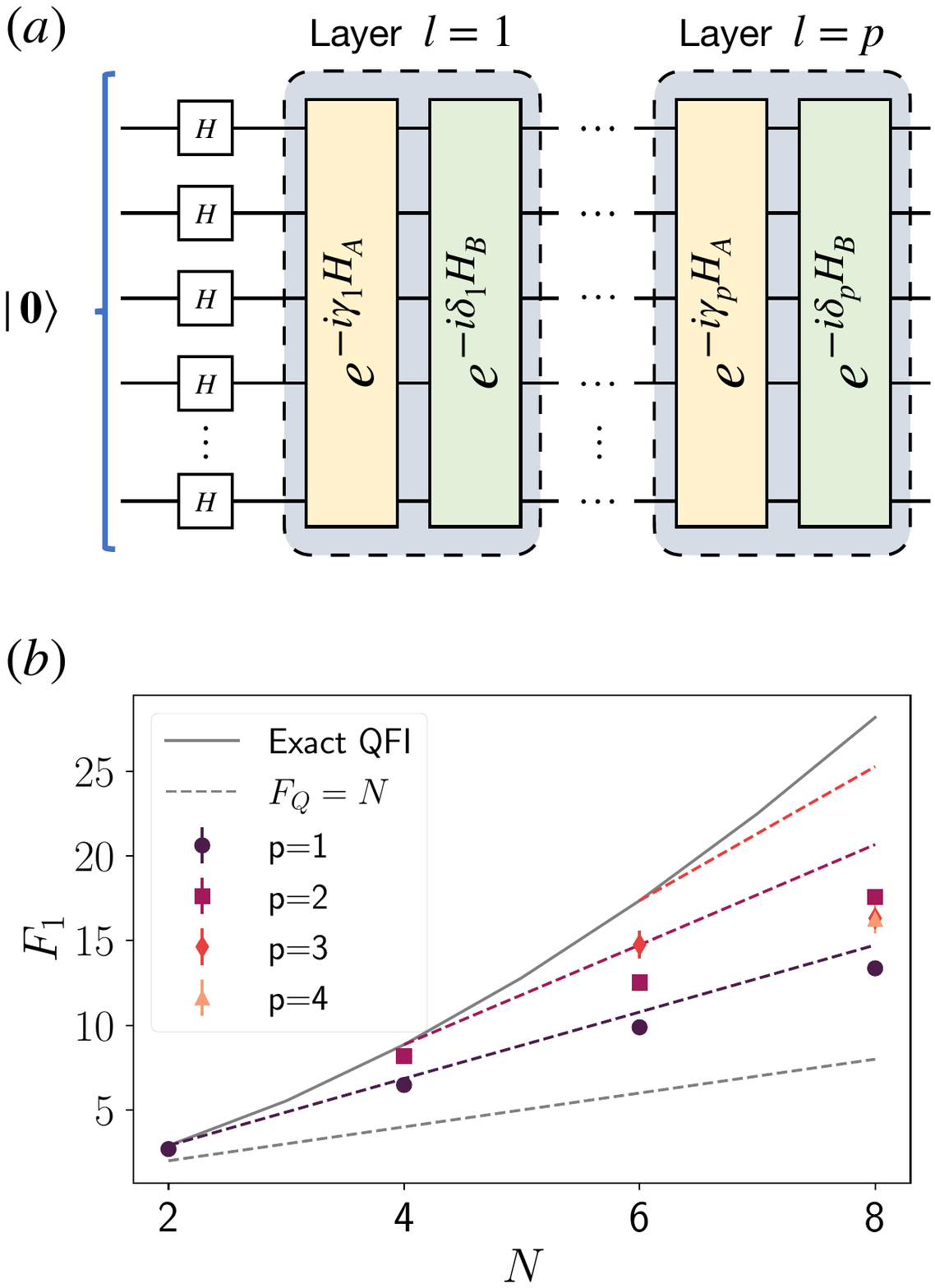}   
    \caption{Experimental results for the lower bounds of the QFI for the ground state of the Ising model at the critical point. $(a)$ Sketch of the circuit used to variationally prepare the ground state. 
    $(b)$ Results for $F_1$ estimated with the robust estimator as a function of the number of qubits $N$, for different circuit depth $p$.  The solid grey line corresponds to the exact QFI value. The colored dashed lines correspond to the theoretical value for the given depth of the circuit $p$. The dashed grey line denotes the threshold $F_Q=N$, above which the state is entangled. The measurement budget employed is described in Sec. \ref{sec:budget}.}
    \label{fig:fig3}
\end{figure}
\vspace{2em}
{\renewcommand\addcontentsline[3]{}\subsection{Ground state of the critical TFIM}\label{sec:QAOA}}
While the GHZ has a simple analytical wavefunction, we find it instructive to apply our protocol now on a quantum state with a more complex multipartite entanglement structure. We study here the behavior of the QFI at a critical point that presents a rich structure of multipartite entanglement~\cite{Smerzi2012,PappalardiQFI2017,gabbrielli_multipartite_2018,frerot2018quantum}.
In particular, we consider the transverse field Ising model (TFIM) Hamiltonian
\begin{equation}\label{eq:TFIM}
    H=-J\sum_j \sigma^z_j \sigma^z_{j+1} -h \sum_j \sigma^x_j,
\end{equation}
where $h$ is the transverse field and we set $J=1$. It displays a quantum phase transition at $h=1$ that manifests as a growth of multipartite entanglement that can be witnessed by the QFI~\cite{hauke_measuring_2016,frerot2018quantum}.
We employ classical numerical simulations for estimating variationally the ground state at the critical point, optimizing the parameters of a circuit as it is done for the quantum adiabatic optimization algorithm~\cite{QAOAreview} (See Fig.~\ref{fig:fig3}$(a)$).
Then we study the interplay between the depth $p$ of the circuit realized and the approximation of the ground state in an actual experiment.
Indeed, in recent times there has been significant interest in measuring the QFI in states prepared through variational circuits on current quantum processors~\cite{Koczor_2020,cerezo_variational_2021,Beckey2022}.

The preparation of the state entails a series of unitary quantum evolutions under the non-commuting terms in Eq.~\eqref{eq:TFIM}, i.e. $H_A = -J\sum_j \sigma^z_j \sigma^z_{j+1}$ and \mbox{$H_B=-h \sum_j \sigma^x_j$}, that are applied to an initial quantum state $\ket{\psi_0}$ (Fig.~\ref{fig:fig3}$(a)$).
The final state after $p$ layers can be written as:
\begin{equation}
    \ket{\psi(\bm{\delta}, \bm{\gamma})}=\prod_{l=1}^p \mathrm{e}^{-i \delta_l H_B} \mathrm{e}^{-i \gamma_l H_A}\ket{\psi_0}
\end{equation}
where  the  `angles' $\delta_l$ and $\gamma_l$ are variational parameters used in the $l$-th layer to minimise the final energy $\bra{\psi(\bm{\delta},\bm{\gamma})}H\ket{\psi(\bm{\delta},\bm{\gamma})}$. 
The optimal parameters are found by employing a suitable optimization algorithm.
In the particular case of our target state, it has been shown that it could be prepared exactly in $p=N/2$ steps, where $N$ is the total number of qubits~\cite{WenWei2019}. 

In Fig.~\ref{fig:fig3}$(b)$ we plot $F_1$ for different values of the depth $p$ of the circuits as a function of the number of qubits $N$ for the robust estimator. 
The solid grey line represents the exact numerical value of the QFI. Instead, the colored dashed lines correspond to the exact value of QFI for different depths $p=1,2,3$ from bottom to top.
Our first observation consists in remarking that we generate and certify the presence of entanglement in all our prepared states as $F_Q\ge F_1\ge N$~\cite{Pezze2018} within error bars for all values of depths $p$ and system size $N$. The corresponding threshold is shown as a dashed grey line in the plot. 

We observe that a larger circuit depth is not tightly linked to a higher measured value of $F_1$.
Indeed, the increase in the circuit depth $p$, incorporates more noisy gates that reduce the fidelity of the prepared state compared to its true target state. 
This results in a decrease of the QFI estimation compromised by the noisy state preparation that is captured very well in Fig.~\ref{fig:fig3}$(b)$. 
In the ideal scenario, increasing the number of layers should guarantee better convergence to the target state. The effect of noise is clearly shown in App.~\ref{moreexpresults}, where we also provide the estimation of the fidelity of the state preparation and the purity of the prepared state.

Importantly, we establish the presence of multipartite entanglement via $F_1$ as we violate the entanglement witness $F_1>\Gamma(N,k=2)$~\cite{Toth2012, Hyllus2012}. This confirms the presence of an entanglement depth of $k=3$ for all prepared states of system-size $N>2$, as the experimental points are above the witness depicted by the dashed dark grey line in Fig.~\ref{fig:fig3}$(b)$.
Thus, our method allows us to quantify the true metrological power in the form of generating multipartite entanglement in our noisy prepared states.
Additionally, we remark here that we have focused on the bound $F_1$ because it provides more reliable estimates with respect to the other estimators ($F_0$ and $F_2$) even if their qualitative behavior is the same. The reason behind this choice is that $F_2 > F_1 > F_0$, but the higher the order, the larger the statistical error at fixed measurement budget, as explained in Sec.~\ref{sec:datacompression}. We provide the other experimental results in App.~\ref{SM:QAOA}.

\vspace{2em}
{\renewcommand\addcontentsline[3]{}\section{Conclusions}}
In this paper, we have provided an experimental estimation of the quantum Fisher information (QFI). This was achieved on a quantum processor with up to 13 qubits based on the measurements of a \emph{converging} series of polynomial lower bounds. By combining advanced methods from the randomized measurement toolbox, we have been able to overcome drifting gate and readout errors and obtain a robust and unbiased estimator for the QFI. 

We applied our method to two different states: GHZ states and the ground state of the TFIM at the critical point.
For the former, our measurements are in perfect agreement with theoretical predictions and allow us to witness the presence of multipartite entanglement. 
With the error mitigation procedure that we introduce here, we observed that all our prepared GHZ states were GME.
In the variational preparation of the ground state of the critical TFIM, we utilize the estimated QFI to observe an interesting trade-off. While the theoretical approximation accuracy of the ground state increases with the circuit depth and is optimal at depth $p=N/2$,  the best estimation of the theoretically predicted ground state QFI is obtained with a smaller circuit depth. We attribute this effect to noise and decoherence increasing with circuit depth as well.

We have gone beyond previous work aiming at estimating 
a converging series of lower bounds on the QFI (rather than a single lower bound to it), employing larger system sizes, and obtaining a better convergence to the true value of the QFI for the prepared state. This has been possible, by exploiting several state-of-the-art protocols under the umbrella of randomized measurements.

We stress that our method is well-suited for following the drifting errors in the hardware as experimentally shown in App.~\ref{SM:batchexp_check}. Performing a calibration at the beginning of the whole experiment is not sufficient for taming and understanding the errors in the randomized measurement protocol, of which we show evidence in App.~\ref{SM:batchvsfull}.

Furthermore, our approach is not limited to the measurement of the QFI. Our results extend easily to obtain robust and unbiased estimators for arbitrary nonlinear multi-copy functionals that can be expressed as observables acting on multiple copies of the quantum state. This extends the applicability of our methodology beyond the QFI and opens up possibilities for other quantum information processing tasks, such as exploring many-body entanglement phases by measuring partial transpose moments~\cite{carrasco2022entanglement}. Additionally, as the robust calibration method is memory efficient, it can be performed to measure observables such as energy estimation of the ground state of quantum chemistry Hamiltonians prepared on large-scale quantum devices~\cite{Hempel_quantum_2018,Huang_efficient_2021}  that can be further boosted by employing common randomized measurements techniques~\cite{vermersch2023enhanced}. Finally, our method could be used in combination with machine-learning approaches to learn complex phases of matter with robust shadows~\cite{Huang_provable_2022,lewis2023improved}.\\

\vspace{2em}
{\renewcommand\addcontentsline[3]{}\section*{Acknowledgements}}
We thank S.~Flammia for his valuable comments on our manuscript.
Work in Grenoble is funded by the French National Research Agency via the JCJC project QRand (ANR-20-CE47-0005), and via the research programs EPIQ (ANR-22-PETQ-0007, Plan France 2030), and QUBITAF (ANR-22-PETQ-0004, Plan France 2030).
B.V.\ acknowledges funding from the Austrian Science Foundation (FWF, P 32597 N).
A.R.\ acknowledges support by Laboratoire d'excellence LANEF in Grenoble (ANR-10-LABX-51-01) and from the Grenoble Nanoscience Foundation. 
A.E.\ acknowledges funding by the German National Academy of Sciences Leopoldina under grant number LPDS 2021-02 and by the Walter Burke Institute for Theoretical Physics at Caltech.
For some of our numerical simulations, we used the quantum toolbox QuTiP~\cite{Johansson2013}. 



\bibliographystyle{apsrev4-1}
{\renewcommand\addcontentsline[3]{}\bibliography{biblio}}

\newpage

\onecolumngrid
\appendix
\appendixpage
\tableofcontents

\section{Converging series of lower bounds of the QFI}\label{SM:Fnappendix}
As shown in Ref.~\cite{Rath_QFI_2021}, the QFI can be expanded in terms of a Taylor series in the eigenvalues $\lambda_\mu$ of the density matrix $\rho = \sum_\mu \lambda_\mu \ketbra{\mu}{\mu}$. This reads as: 
\begin{equation}
    F_Q =  2   \sum_{\ell = 0}^{\infty}\sum_{(\mu,\nu), \lambda_\mu + \lambda_\nu > 0}(\lambda_\mu - \lambda_\nu)^2 (1-\lambda_\mu - \lambda_\nu)^\ell |\bra{\mu}A\ket{\nu}|^2.
\end{equation} 
We note that each term in the infinite sum is positive.
Truncating the summation at a finite value $n$, we thus obtain a converging series of polynomial lower bounds $F_n$ that can be measured experimentally:
\begin{equation}
    F_n =  2\sum_{\ell = 0}^{n} \sum_{(\mu,\nu), \lambda_\mu + \lambda_\nu > 0} (\lambda_\mu - \lambda_\nu)^2 (1-\lambda_\mu - \lambda_\nu)^\ell |\bra{\mu}A\ket{\nu}|^2 = 2 \sum_{q = 0}^{n} \binom{n+1}{q+1} (-1)^q \sum_{m = 0}^{q+2} C_{m}^{(q)} \Tr(\rho^{q+2-m}A\rho^{m} A), \label{eq:QFI-hockey}
\end{equation}
where we have introduced the coefficients $C_{m}^{(q)} = \binom{q}{m} -2\binom{q}{m-1}+\binom{q}{m-2}$, with the binomial coefficients defined such that $\binom{q}{m'} = 0$ if $m' < 0$ or $m' > q$.
The last equality can be proven by injecting the eigenvalue decomposition of $\rho$ in the right-hand side and rearranging the sums~\cite{Rath_QFI_2021}.

\section{Quantum properties under global deporalization}\label{sec:depoQFI}
Let us consider a quantum state defined as $\rho(p_D) = (1-p_D)\ketbra{\psi}{\psi} + p_D \mathbb{1}/2^N$, where $\ket{\psi}$ is a pure state and $\mathbb{1}/2^N$ is the fully mixed state.
The state $\rho(p_D)$ is mixed with global depolarizing noise of strength $p_D$. 
The distinct eigenvalues of $\rho(p_D)$ are $\lambda_1 = (1-p_D)+p_D/2^N$ (with multiplicity 1) and $\lambda_2 = p_D/2^N$ (with multiplicity $2^N-1$).
As shown previously in~\cite{Hyllus2012}, for $\rho(p_D)$, the QFI is given (replacing the eigenvalues in Eq.~\eqref{eq:qfi}) by
\begin{equation}
    F_Q = 4 \left( \bra{\psi}A^2\ket{\psi} - \bra{\psi}A\ket{\psi}^2\right)\frac{(1-p_D)^2}{1-p_D+2p_D/2^N}. \label{eq:qfidepo}
\end{equation}
Similarly, we notice that for this specific state, all non-zero terms 
in Eq.~\eqref{eq:QFI-hockey} are equivalent to $(1-p_D)^2(p_D - 2p_D/2^N)^\ell$. Thus we can provide an expression of the lower bounds $F_n$ under global depolarization noise as:
\begin{equation}
    F_n =  4 \left( \bra{\psi}A^2\ket{\psi} - \bra{\psi}A\ket{\psi}^2\right)  (1-p_D)^2 \sum_{\ell = 0}^n (p_D - 2p_D/2^N)^\ell 
\end{equation}
Analogously, we can also express the analytical form of the purity for the state $\rho(p_D)$ as
\begin{equation}
    \Tr(\rho(p_D)^2) = (1-p_D)^2 + \frac{2p_D - p_D^2}{2^N}.
\end{equation}
Assuming this specific noise model, one can perform multiple interesting analyses with the above analytical expressions. 
One such investigation is to better understand the convergence of the lower bounds $F_n$ to a finite value of QFI as a function of the bound order $n$.
For this purpose, we could use the experimentally recorded values of purity to invert the above purity expression and extract values of $p_D$ for a given state of interest.
We perform this analysis for the 8 and 10 qubit GHZ states as shown in Fig.~\ref{fig:fig2}$(a)$. 
The mitigated values of purities as estimated and shown in Fig.~\ref{fig:fig2}$(a)$ are respectively $0.89 \pm 0.0066$ and $0.86 \pm 0.017$ which give the corresponding values of $p_D$ as $0.056 \pm 0.007$ and $0.072 \pm 0.018$ for the 8 and 10 qubit GHZ states.
One can estimate easily the theoretical value of the QFI using Eq.~\eqref{eq:qfidepo} under this noise assumption.
This presents an easy method to check the convergence of the lower bounds to a finite value of QFI as presented in Fig.~\ref{fig:fig2} of the main text.
\section{Derivation of the robust shadow estimator with local noise}\label{SM:robust}

In this section we construct the robust classical shadow estimator given in Eq.~\eqref{eq:robshadows}, equivalent to the one presented for the first time in Ref.~\cite{Chen2020}. We consider  a situation where we have performed randomized measurements on a $N$-qubit state $\rho$, which are affected by noise.  We assume that the noise is gate-independent, Markovian and stationary within each iteration, and that it occurs between the random unitaries and the measurements (not before the unitaries). This ensures that we can model  noisy  randomized measurements as $\mathcal{M} \circ\Lambda^{(i)} \circ \mathcal{U}^{(r)}$ where $\mathcal{U}^{(r)}$ is the ideal unitary channel describing the application of ideal random unitaries $U^{(r)}$, $\Lambda^{(i)}$ is the noise channel in iteration $i$, encapturing gate noise and  readout errors,  and $\mathcal{M}$ is the measurement channel, describing an ideal computational basis measurement~\cite{Chen2020}. In addition, we assume local noise, i.e.\ the noise channel decomposes as $\Lambda^{(i)}=\Lambda^{(i)}_1\otimes \dots \otimes \Lambda^{(i)}_N$ and local random unitaries, i.e.\ the ideal unitary channel is realized with local unitary transformations $U^{(r)}=U^{(r)}_1\otimes \dots \otimes U^{(r)}_N$. Here, the local unitaries $U_j^{(r)}$ are sampled independently and uniformly from the circular unitary ensemble, i.e.\ the Haar measure on the unitary group $U(2)$.

As described in the main text, we employ first a calibration protocol, equivalent to the one described in Ref.~\cite{Chen2020}, to characterize the local noise channel  $\Lambda^{(i)}$  in terms of $N$ parameters $G^{(i)}_j$. To perform this calibration, we assume that the state $\ket{\0} = \ket{0}^{\otimes N}$ can be prepared with a high fidelity in our experiment. The calibration results are then used to build an unbiased estimator $\hat{\rho}$ of the density matrix $\rho$ -- a robust classical shadow -- from randomized measurements performed on $\rho$, that mitigates the noise errors induced by $\Lambda^{(i)}$. 

In the remainder of this section, we drop the superscript $i$ denoting the iteration of the experiment to simplify the notation.

\subsection{Robust shadow from randomized measurements}\label{SM:rob_from_meas}
Under the noise assumptions described above, noisy randomized measurements provide access to the probability distribution of the measured bit-strings $\s = (s_1, \dots, s_N)$, conditioned on the application of a local random unitary $U^{(r)} = U_1^{(r)} \otimes \dots \otimes U_N^{(r)}$:
\begin{equation}
    P_\Lambda(\s|U^{(r)}) = \bra{\s} \Lambda(U^{(r)}\rho\, {U^{(r)}}^\dag ) \ket{\s}
    =\Tr(\rho\, {U^{(r)}}^\dag \Lambda^*( \ketbra{\s}{\s})U^{(r)}),
\end{equation}
where $\Lambda$ is the trace-preserving noise channel and $\Lambda^*$ is its adjoint.
We aim to construct an unbiased estimator of $\rho$ -- robust classical shadow -- in terms of the statistics of $P_\Lambda(\s|U^{(r)})$. 
We choose an ansatz of the form
\begin{equation}
\label{eq:ansatz}
\begin{aligned}
    \tilde{\rho}^{(r)}&=\sum_{\s} P_\Lambda(\s|U^{(r)}) \, {U^{(r)}}^{\dag} O(\s) U^{(r)}\\
    &=\sum_{\s} \Tr(\rho\, {U^{(r)}}^{\dagger}\Lambda^*(\ketbra{\s}{\s})U^{(r)})\, {U^{(r)}}^{\dag}O(\s)U^{(r)} \\
    &=\sum_{\s}\Tr_1\left( \left( \rho \otimes \mathbb{1} \right) ({U^{(r)}}^{\dag})^{\otimes 2} \big[ \Lambda^*(\ketbra{\s}{\s})\otimes O(\s) \big] {U^{(r)}}^{\otimes 2} \right)
    \end{aligned}
\end{equation}
with $O(\s) = \bigotimes_j O_j(s_j)$ being a local hermitian operator, which we take to be diagonal in the computational basis, and $\Tr_1$ denoting the partial trace over the first copy of the $N$-qubit system.
The ensemble average over the random unitaries $U^{(r)}$ yields 
\begin{equation} \label{eq:E_tilde_rho_r}
    \mathbb{E}[\tilde{\rho}^{(r)}]=\Tr_1\left[(\rho\otimes \mathbb{1})\Phi^{(2)}\left(\sum_{\s}  \Lambda^*( \ketbra{\s}{\s})
    \otimes O(\s) \right)\right]=\Tr_1\left[(\rho\otimes \mathbb{1})\Phi^{(2)}(Q)\right]
\end{equation}
with 
\begin{align}
Q = \sum_{\s}  \Lambda^*( \ketbra{\s}{\s})
    \otimes O(\s)  = \bigotimes_{j =1}^N \left[\sum_{s_j = 0,1}  \Lambda_j^*( \ketbra{s_j}{s_j})
    \otimes O_j(s_j) \right] = \bigotimes_{j=1}^N Q_j.
\end{align}
Here, we used the local noise assumption (noting that $\Lambda^* = (\bigotimes_j \Lambda_j)^* = \bigotimes_j \Lambda_j^*$) and $\Phi^{(2)}(\cdot) = \mathbb{E}[({U^{(r)}}^{\dag})^{\otimes 2} (\cdot) {U^{(r)}}^{\otimes 2}]$ denotes the two-copy local unitary `twirling channel'~\cite{Elben2018a}. It evaluates to
\begin{equation}\label{eq:twirlingformula}
\Phi^{(2)}(Q) = \left(\frac{1}{3}\right)^N \bigotimes_{j =1}^N
\left( \left(\mathrm{Tr}(Q_j)-\frac{1}{2}\mathrm{Tr}(\S_j Q_j)\right) \mathbb{1}^{(2)}_j +\left(\mathrm{Tr}(\S_j Q_j)-\frac{1}{2} \mathrm{Tr}(Q_j)\right)\S_j\right)
\end{equation}
with the swap operator  \mbox{$\S_j=\sum_{s_{j_1}, s_{j_2}}\ketbra{s_{j_2}}{s_{j_1}} \otimes \ketbra{s_{j_1}}{s_{j_2}}$} acting on two copies of qubit $j$ and $\mathbb{1}^{(2)}_j =\mathbb{1}_j \otimes \mathbb{1}_j$ the identity.

The estimator $\tilde{\rho}^{(r)}$ is an unbiased estimator of $\rho$ if the average over the Haar random unitaries yields the true density matrix of the quantum state, $\mathbb{E}\left[\tilde{\rho}^{(r)}\right]=\rho$. Observing that $\Tr_1\left(\left(\rho \otimes \mathbb{1}\right)\S\right)=\rho$ where $\S=\bigotimes_{j=1}^N \S_j$ is the swap operator between two copies of the entire system, we thus find, from Eq.~\eqref{eq:E_tilde_rho_r}, that the estimator is unbiased for any state $\rho$ if and only if $\Phi^{(2)}(Q) = \S$, or equivalently, using Eq.~\eqref{eq:twirlingformula},
\begin{align}
\left(\mathrm{Tr}(Q_j)-\frac{1}{2}\mathrm{Tr}(\S_j Q_j)\right) \mathbb{1}^{(2)}_j +
\left(\mathrm{Tr}(\S_j Q_j)-\frac{1}{2} \mathrm{Tr}(Q_j)\right) \S_j =3 \S_j\quad \forall j. \label{eq:phiSwap}
\end{align}
On top of the assumption that  $O_j(s_j)$ is diagonal in the computational basis, we further assume that it is of the form \mbox{$O_j(s_j)=\alpha_j\ketbra{s_j}{s_j}+\beta_j\mathbb{1}$},  with $\alpha_j$, $\beta_j$ real numbers that do not depend on $s_j$. With this, we can evaluate the terms appearing in Eq.~\eqref{eq:phiSwap} above as follows:
\begin{align}
\Tr(Q_j) &=\sum_{s_j} \Tr(\Lambda^*(\ketbra{s_j}{s_j}))\Tr(O_j(s_j))
= \sum_{s_j} \Tr(\Lambda^*(\ketbra{s_j}{s_j}))(\alpha_j+2\beta_j) \notag
\\
&= \Tr(\Lambda(\mathbb{1}))(\alpha_j+2\beta_j)
=2\alpha_j+4\beta_j
,\\
\Tr(\S_j Q_j) &= \sum_{s_j}\Tr(\Lambda^*(\ketbra{s_j}{s_j})O_j(s_j)) \notag \\
&=\alpha_j \sum_j \Tr(\Lambda^*(\ketbra{s_j}{s_j})\ketbra{s_j}{s_j})+2\beta_j
=2\alpha_j G_j+2\beta_j,
\end{align}
where we have used also that the noise channel is trace preserving and $\Tr(\S A \otimes  B)=\Tr(AB)$.
Here, we have introduced the quantity
\begin{equation}\label{SM:Gj}
G_j
=
\frac{1}{2}\sum_{s_j}\bra{s_j} \Lambda^{}_{j}(\ketbra{s_j}{s_j}) \ket{s_j},
\end{equation}
that contains all the relevant information on the noise acting on qubit $j$ during the randomized measurement protocol, and which we interpret as the average `survival probability' of the two basis states of qubit $j$.
Thus, to characterise the noise that affects the experimental protocol we only need to learn how it acts on the computational basis states $\ket{s_j}$. 
With the above expressions, inverting Eq.~\eqref{eq:phiSwap} gives
\begin{equation}\label{eq:localajbj}
  \alpha_j=\frac{3}{2 G_j-1}, \qquad
  \beta_j= \frac{G_j-2}{2 G_j-1}.
\end{equation}
Inserting this into Eq.~\eqref{eq:ansatz}, we finally can write the estimator $\tilde{\rho}$ as 
\begin{equation}\label{eq:robustSM}
\begin{aligned}
 \tilde{\rho}^{(r)}&=\sum_{\s} P_\Lambda(\s|U^{(r)}) \bigotimes_{j=1}^N \left( \alpha_j {U^{(r)}_j}^{\dag} \ketbra{s_j}{s_j}U^{(r)}_j +\beta_j \mathbb{1}\right) \\
 &=\sum_{\s} P_\Lambda(\s|U^{(r)})\bigotimes_{j=1}^N \left( \frac{3}{2 G_j-1} {U^{(r)}_j}^{\dag} \ketbra{s_j}{s_j} U^{(r)}_j+\frac{G_j-2}{2 G_j-1} \mathbb{1}\right).
\end{aligned}
\end{equation}
In the absence of noise $G_j=1$, $\forall j$, so that the usual formula for the estimator of the density matrix from randomized measurements is recovered: $O_j(s_j)=3\ketbra{s_j}{s_j}-\mathbb{1}$~\cite{Elben2019,Huang2020}.  For a fully depolarising channel, on the other hand, one gets $G_j=1/2$, in which case we are not able to extract any information by measuring the state as the coefficients in Eq.~\eqref{eq:localajbj} diverge.

\subsection{Calibration step}\label{SM:calibration}
The parameters in Eq.~\eqref{eq:localajbj} rely on the estimation of $G_j$.
It is based on the calibration procedure described in the main text (Sec.~\ref{sec:EP}). In a nutshell, the system is prepared in a state with high fidelity, namely $\ket{\mathbf{0}} \equiv \ket{0}^{\otimes N}$, and the randomized measurement protocol is applied.
We show here that $G_j$ can be directly linked to the random unitaries \mbox{$U^{(r)} = U_1^{(r)} \otimes \dots \otimes U_N^{(r)}$}, with \mbox{$r= 1, \dots, N_U$}, and the bit-strings of measurement outcomes \mbox{$\s^{(r,m)} = (s^{(r,m)}_1,\dots , s^{(r,m)}_N)$} with \mbox{$m = 1, \dots, N_M$}.

Let us introduce the following quantity
\begin{equation}\label{eq:Cj}
  C_j=\sum_{s_j=0,1} \mathbb{E}\left[\bra{s_j} \Lambda_j(U^{(r)}_j\ketbra{0}{0} {U^{(r)}_j}^{\dag} ) \ketbra{s_j}{s_j} U^{(r)}_j \ketbra{0}{0} {U^{(r)}_j}^{\dag}  \ket{s_j}\right],
\end{equation}
where $\ket{0}$ represents the calibration state of the single qubit $j$ and $\mathbb{E}[\cdot]$ is the average over the circular unitary ensemble. We can define an estimator as:
\begin{equation}\label{eq:estimatorCjSM}
\hat{C}_j =\frac{1}{N_U}\sum_{r} \sum_{s_j=0,1} \hat{P}(s_j|U_j^{(r)}) P(s_j|U_j^{(r)}),
\end{equation}
where $\hat{P}(s_j|U_j^{(r)}) = \sum_{m = 1}^{N_M} \frac{\delta_{s_j, s_j^{(r,m)}}}{N_M}$ is the estimated (noisy) Born probability and $P(s_j|U_j^{(r_i)}) = |\bra{s_j} U^{(r)}_j\ket{0}|^2$ is the theoretical (noiseless) one. The information on the noise is contained in $\hat{P}(s_j|U_j^{(r)})$, that approaches the theoretical noisy Born probability $P_\Lambda(s_j|U_j^{(r)})$ in the limit $N_M\rightarrow \infty$.
Thus, since with our noise model $\mathbb{E}_{\rm QM}[\hat{P}(s_j|U^{(r)})]=\bra{s_j} \Lambda_j(U^{(r)}_j\ketbra{0}{0} {U^{(r)}_j}^{\dag} ) \ket{s_j}$, we have $\mathbb{E}\big[\mathbb{E}_{\rm QM}[\hat{C}_j]\big]= C_j$, i.e., $\hat{C_j}$ is an unbiased estimator for $C_j$. Here $\mathbb{E}_{\rm QM}[\cdot]$ is the quantum mechanical average over the Born probabilities. 
We note that under the (idealized) assumption of strictly gate-independent noise (same noise channel $\Lambda_j$ for any $U^{(r)}_j$, including the idle gate $\mathbb{1}_j$), we could measure $G_j$ directly from its definition, Eq.~\eqref{SM:Gj}. In practice, we expect that  $C_j$ (and its estimator $\hat{C}_j$)  captures the actual noise acting during the measurement stage more faithfully, as exactly the same experimental resources are employed and any weakly gate dependent noise is averaged (twirled) to yield approximately the same gate-independent average noise channel, $\mathbb{E}[\Lambda_U(\rho)]\approx \Lambda(\mathbb{E}[U\rho U ^\dag])$. We refer to more details on gate-dependent noise in Ref.~\cite{Chen2020}.

Let us now link $C_j$ and the quantity $G_j$ in Eq.~\eqref{SM:Gj}.
We observe that $C_j$ can be written as
\begin{equation}
C_j=\mathbb{E}\left[\bra{0}^{\otimes 2} {{U^{(r)}_j}^{\dag}}^{\otimes 2}\left(\sum_{s_j=0,1}\Lambda_j^{*}\left(\ketbra{s_j}{s_j}\right)\otimes \ketbra{s_j}{s_j}\right){U^{(r)}_j}^{\otimes 2}\ket{0}^{\otimes 2}\right], 
\end{equation}
where we have used the property $\mathrm{Tr}(\Lambda(A)B)=\mathrm{Tr}(A\Lambda^*(B))$.
As in the previous section, we can express the average in $C_j$ over unitaries  in terms of a twirling channel  $\Phi_{j}^{(2)}(Q_j)$ (a single-qubit version of the 2-copy channel $\Phi$ introduced before).
In this case we write the two-copy operator $Q=\bigotimes_{j=1}^N Q_j$ with $Q_j=\sum_{s_j=0,1}\Lambda_j^{*}\left( \ketbra{s_j}{s_j}\right)\otimes \ketbra{s_j}{s_j}$.
Using again the twirling formula in Eq.~\eqref{eq:twirlingformula} (now for two copies of a single qubit) one obtains
\begin{equation}\label{eq:twirlC}
     C_j=
 \bra{0}^{\otimes 2}
 \Phi_j^{(2)}\left(\sum_{s_j}\Lambda^*(\ketbra{s_j}{s_j})\otimes \ketbra{s_j}{s_j}\right)  \ket{0}^{\otimes 2}=
 \frac{1+G_j}{3}.
\end{equation}
The link between $C_j$ and $G_j$ being clear, one can define an estimator for $G_j$ in terms of the one for $C_j$ in Eq.~\eqref{eq:estimatorCjSM}:
\begin{equation}\label{eq:Gestimator}
    \hat{G}_j=3\; \hat{C}_j-1.
\end{equation}
In the absence of noise one can check that $G_j=1$ and $C_j=\frac{2}{3}$, $\forall j=1,\dots,N$~\cite{Huang2020}.
We remark here that all our results are compatible with the ones in Ref.~\cite{Chen2020}, where a slightly different formalism has been employed.

We have discussed in Sec.~\ref{subsec:crm} that the statistical accuracy of  the estimator of $G_j$ can be improved by making use of \emph{common randomized numbers}~\cite{vermersch2023enhanced} to define an estimator with smaller variance with respect to $\hat{G}_j$.
 
The enhanced estimator $\hat{\mathcal{G}}_j$ is connected as well to the quantity $C_j$ that we access in the experiments, as in Eq.~\eqref{eq:CCRestimator}. Explicitly, we write:
\begin{equation}
    \hat{\mathcal{G}}_j=3\;\hat{C}_j-1-B_j + \mathbb{E}[B_j], 
\end{equation}
where $B_j$ has been defined in Eq.~\eqref{eq:Bmaintext}.
Therefore the common randomized measurements procedure enters solely into the post-processing.
The variance of estimators obtained from such practice has been studied analytically in Ref.~\cite{vermersch2023enhanced}, where it was shown that involving positively correlated random variables, as above, indeed allows one to significantly reduce the variance upper bounds.
In the following section we compare based on our experimental data this estimator with the one introduced in Eq.~\eqref{eq:Gestimator}.

\subsection{Experimental comparison of the estimators \texorpdfstring
{$\hat{G}_j$}{} and \texorpdfstring{$\hat{\mathcal{G}}_j$}{}}\label{SM:comparisonG}
\begin{figure}
    \centering
    \includegraphics[width=0.5\linewidth]{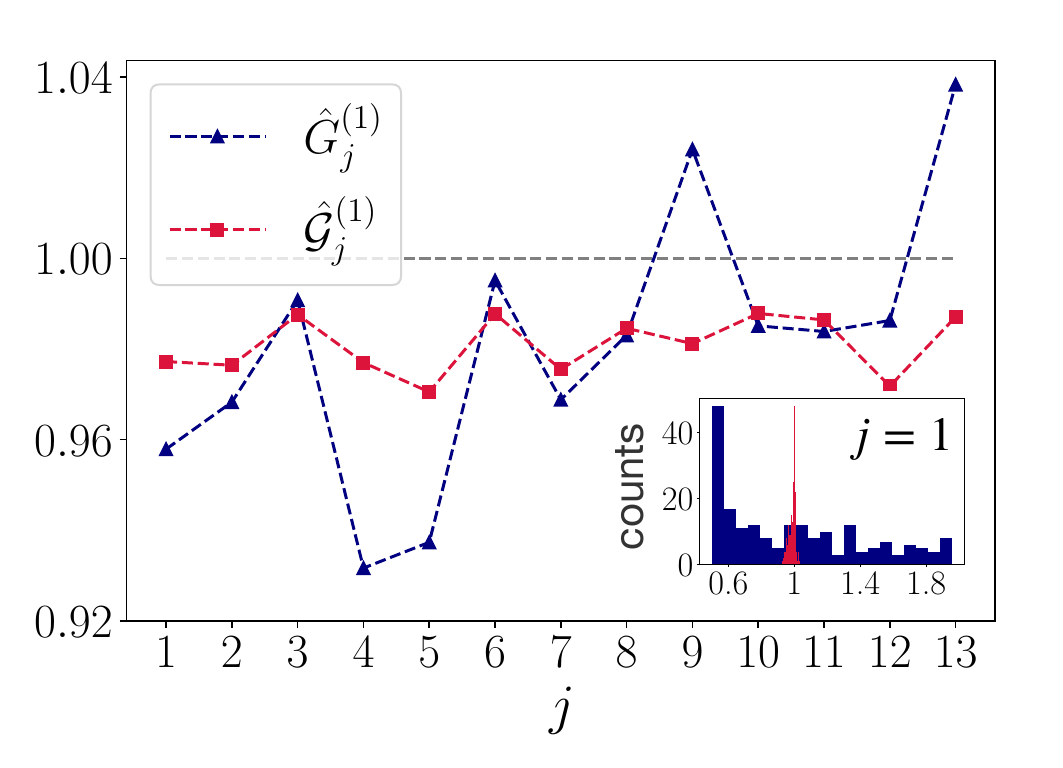}
    \caption{Comparison of $G_j$ using the enhanced estimator $\hat{\mathcal{G}}_j$ and the previous estimator $\hat{G}_j$ for a 13-qubit state on `ibm\_prague'. The quantities are measured according to the calibration protocol described in Sec.~\ref{sec:EP} and depicted in Fig.~\ref{fig:fig1}. $N_U^{\mathrm{tot}}= N_I N_U = 27000 $ (number of unitaries in the randomized measurement protocol) and $N_M=1000$ (number of measurements per unitary). In the inset we compare the estimators for iteration $i=1$, i.e. $\hat{G}^{(1)}_j$ (blue) and $\hat{\mathcal{G}}^{(1)}_j$ (red), for the first qubit ($j=1$) by plotting a histogram where each occurrence corresponds to an element of the sum over $r$ in Eq.~\eqref{eq:estimatorCjSM} for $\hat{G}^{(1)}_j$, and of the analogous sum for $\hat{\mathcal{G}}^{(1)}_j$.}
    \label{fig:fig4}
\end{figure}

Let us consider the $N=13$ qubit experiment that has been performed on the `ibm\_prague' processor.  We have performed a calibration of the device as described in Sec.~\ref{sec:EP} and depicted in the step (i) of  Fig.~\ref{fig:fig1}. For each iteration $i=1,\dots,N_I$ and for each applied unitary $U^{(r_i)}$ ($r_i=1,\dots,N_U$), we collect $N_M=1000$ bit-strings of measurement outcomes. From the unitaries and the bit-strings we compute the quantities $\hat{G}^{(i)}_j$ and $\hat{\mathcal{G}}^{(i)}_j$ as defined above, which contain the information about the local errors in the measurement protocol within each iteration $i$.

In Fig.~\ref{fig:fig4} we show the comparison of the two estimators, for iteration $i=1$ and all the qubits ($j=1,\dots,13$). In the inset we show a comparison of the histograms of he values that build up $\hat{G}^{(1)}_j$ and $\hat{\mathcal{G}}^{(1)}_j$ for the first qubit ($j=1$), where each point corresponds to an element of the sum over $r$ in Eq.~\eqref{eq:estimatorCjSM} for $\hat{G}^{(1)}_j$, and of the analogous sum for $\hat{\mathcal{G}}^{(1)}_j$. Remarkably, we observe that the contributions to $\hat{\mathcal{G}}_1^{(1)}$ are much less spread than those of $\hat{G}^{(1)}_1$; in particular the contributions to $\hat{G}^{(1)}_1$ range in $\sim (0.6,1.8)$, while the $\hat{\mathcal{G}}_1^{(1)}$ counts are sharply peaked around $\sim 1$.  We argue this is due to the trick of common random numbers~\cite{vermersch2023enhanced} employed to define $\hat{\mathcal{G}}_j^{(i)}$, which in general allows to reduce the variance of the estimator. The same holds for any qubit $j$.

\section{Experimental results on the noise} \label{SM:experimental_checks}
In this section we perform an experimental analysis on the noise in the quantum platform we employ. We investigate the time dependence of the noise, noticing huge fluctuations in the quantities we use to estimate the errors, and we observe that the most important contribution to the single qubit error can be identified to be caused due to readout errors.

\subsection{Verification of the time dependence of the noise in `ibm\_prague'} \label{SM:batchexp_check}

\begin{figure}
    \centering
    \includegraphics[width=0.5\linewidth]{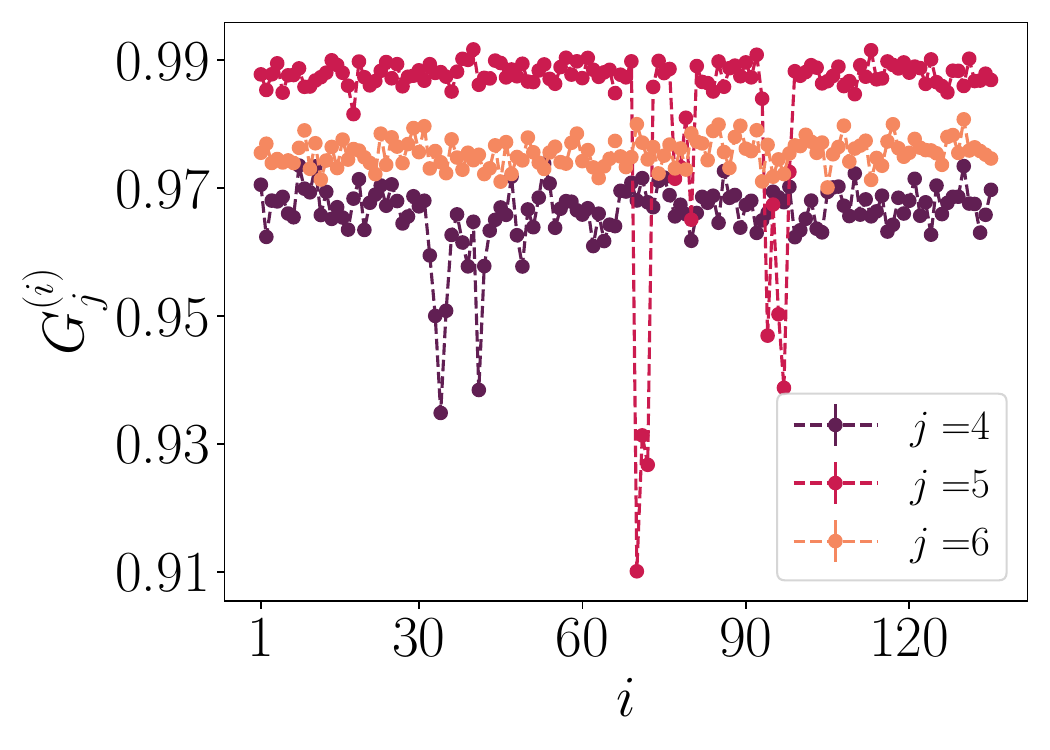}
    \caption{$G_j^{(i)}$  as a function of the iteration $i$ for a 13-qubit state on `ibm\_prague'. The quantity is estimated from the calibration data of the protocol depicted in Fig.~\ref{fig:fig1}, using the estimator $\hat{\mathcal{G}}^{(i)}_j$ of Eq.~\eqref{eq:estimatorMT}. Here $N_U^{\mathrm{tot}}= N_I N_U = 27000 $ (number of unitaries in the randomized measurement protocol) and $N_M=1000$ (number of measurements per unitary). We present the result for qubits $j=4,5,6$.}
    \label{fig:fig5}
\end{figure}

Let us consider again the $N=13$ qubit experiment that has been performed on the `ibm\_prague' processor.  
In Fig.~\ref{fig:fig5} we study the behaviour of $G_j^{(i)}$ (estimated through $\hat{\mathcal{G}}_j^{(i)}$ of Eq.~\eqref{eq:estimatorMT}) as a function of the iterations $i$. The error in the quantum device fluctuates in time, we want to verify that it is important to perform consecutive iterations of experiments to account for the temporal variations in gate and readout errors instead of performing a single calibration in advance.
We plot $G_j^{(i)}$ as a function of $i$ for three different qubits, labeled by $j$. For $j=4,5$, we observe fluctuating events given $G_j^{(i)}$ as a function of the iterations $i$, hinting that it is important to follow the temporal fluctuations of the noise to provide reliable and robust estimations.
This is not the case for all the qubits; e.g., we do not see such fluctuations for $j=6$ in the plot.
Similar effects have been observed in other type of error mitigation protocols with superconducting qubits~\cite{Baheri_2022, hirasaki2023detection}.

\subsection{Check on the origin of the noise}\label{app:originnoise}
Our aim here is to study what is the most important source of errors in the randomized measurement protocol.
\begin{figure}
    \centering
    \includegraphics[width=0.5\linewidth]{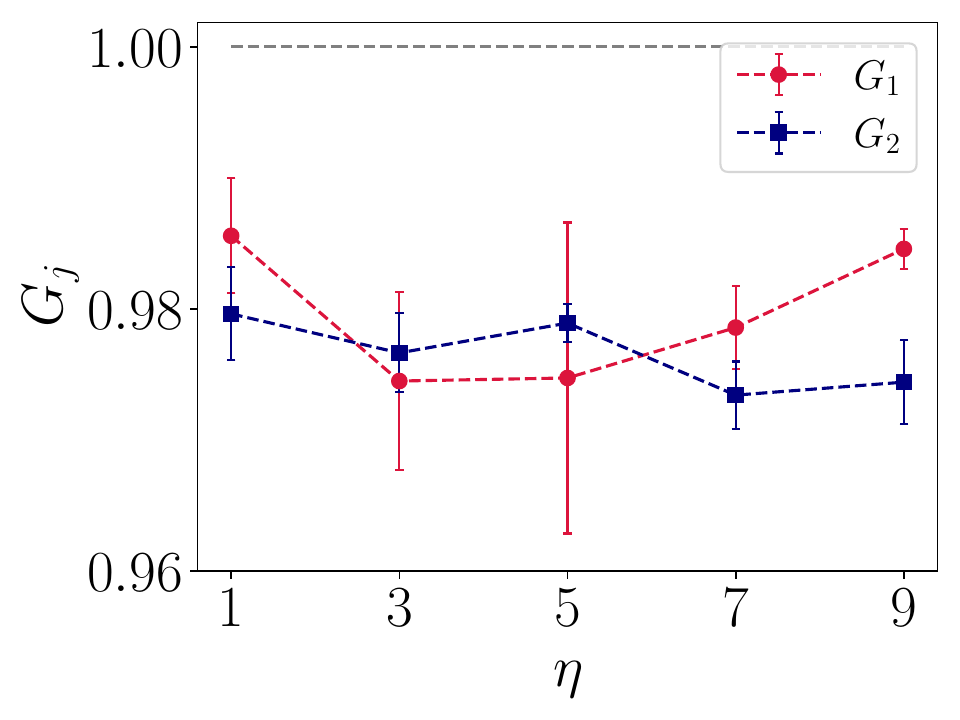}
    \caption{$G_j$ for a two-qubit system realized on `ibm\_lagos', as a function of the number of layers of unitaries $\eta$ applied to the initial state. We use the estimator $\hat{\mathcal{G}}_{j}$ of Eq.~\eqref{eq:estimatorMT}. A value compatible with 1 means that the noise can be neglected. Here $N_U= 800$ (number of unitaries in the randomized measurement protocol) and $N_M=1000$ (number of measurements per unitary).}
    \label{fig6}
\end{figure}
In Fig.~\ref{fig6}
we consider a two-qubit system realized on the `ibm\_lagos' processor. We employ the calibration method described in Sec.~\ref{sec:EP} and depicted in Fig.~\ref{fig:fig1}.
In order to discriminate between the various sources of noises, instead of applying a single unitary $U=\bigotimes_{j=1}^N U_j$, we employ several layers of unitaries, given by the number $\eta$ that are sampled independently and uniformly from the circular unitary ensemble. We measure the quantity $G_j$ as a function of the parameter $\eta$\, using the enhanced estimator $\hat{\mathcal{G}}_j$ of Eq.~\eqref{eq:estimatorMT}.  
The rationale behind this approach is the following. We can write the noise channel $\Lambda$ acting during the measurement protocol as two separate contributions: one due to errors on the unitaries $\Lambda_U$ and one due to the readout $\Lambda_{\textrm{meas}}$ with $\Lambda=\Lambda_{\textrm{meas}} \circ \Lambda_U$. By applying $\eta$ layers of unitaries one would get $\Lambda(\eta)=\Lambda_{\textrm{meas}}\circ (\Lambda_U)^{\eta}$. Following the effect of the noise as a function of $\eta$, we may be able to discriminate the contributions of $\Lambda_U$ and $\Lambda_{\textrm{meas}}$.
This idea can be formalized based on a simple noise model defined by
\begin{equation}
\begin{aligned}
&\Lambda_U(\rho_j)=(1-p_U)\rho_j+\frac{p_U}{3}\sum_{\alpha}\sigma_j^{\alpha}\rho_j\sigma_j^{\alpha},\\
&\Lambda_{\textrm{meas}}(\rho_j)=(1-p_{\textrm{meas}})\rho_j+p_{\textrm{meas}}\sigma_j^{x}\rho_j\sigma_j^{x}.\\
\end{aligned}
\end{equation}
Here $\sigma^{\alpha}=\sigma^x,\sigma^y,\sigma^z$ are single qubit Pauli matrices with $\rho_j$ being a single qubit density matrix. The action of the unitary gates is modeled as a depolarizing noise channel $\Lambda_U$ with parameter $p_U$, while the readout errors are described  by bit flips that happen with probability $p_{\textrm{meas}}$.
The full channel $\Lambda(\eta)(\rho_j)$ applied on a single qubit state $\rho_j$ gives
\begin{equation}
\Lambda(\eta)(\rho_j)
=
(1-\eta p_U-p_{\textrm{meas}})\rho_j
+
\frac{p_U\eta }{3}
\sum_{\alpha}\sigma_j^{\alpha}\rho_j\sigma_j^{\alpha}
+p_{\textrm{meas}}\sigma_j^{x}\rho_j\sigma_j^{x}.
\end{equation}
We can compute explicitly the behavior of $G_j$ at first order in $p_U,p_{\textrm{meas}}\ll 1$ and obtain
\begin{equation}
    G_j(\eta)=1-\frac{2 p_U}{3}\eta -p_\textrm{meas}.
\end{equation}

We observe that the unitary contribution would monotonically decrease $G_j$ as a function of $\eta$ while the readout error yields a fixed shift by $p_{\textrm{meas}}$.
From Fig.~\ref{fig6}, we observe that $G_j$ remains essentially constant within error bars for different values of $\eta$, hence increasing the number of unitaries does not induce more noise (in terms of the parameter $G_j$) in the system. This suggests that the most relevant contribution to the noise in the randomized measurement protocol is due to readout errors.

\section{Verification of the validity of the assumption of local noise}\label{SM:locality}
In this section we propose a method to test the assumption of a local noise channel, i.e. $\Lambda =\bigotimes_{j=1}^{N} \Lambda_j$, that is based on analysing the statistical correlations among qubit pairs. We employ the calibration data that has been used for the mitigation of the QFI results on the prepared GHZ states. 
The section is structure as follows: At first we drop the assumption of locality, i.e. we consider a general noise channel $\Lambda$, and introduce a quantity $\tilde{R}$ that can be used for testing its locality; then, we provide an illustrative analytical example in the case of cross-talk errors for two qubits; finally, we show an experimental indication of the validity of the assumption of locality.

\subsection{Derivation of the estimator of locality of noise}\label{sec:derivationlocality}

Let us start by extending Eq.~\eqref{eq:Cj} to measurements that act on the whole device, writing
\begin{equation} \label{eq:tilde_Cj}
\begin{aligned}
  \tilde C_j&= \mathbb{E}\left[
  \sum_{s_j}\Tr [\bra{s_j} \Lambda(U^{(r)}\ketbra{\0}{{\0}} {U^{(r)}}^\dag ) \ket{s_j}] P(s_j|U_j^{(r)})\right]
  \\
  &=
  \mathbb{E}\left[\sum_{s_j}\Tr[\ketbra{s_j}{s_j} \Tr_{k\neq j}(\Lambda(U^{(r)}\ketbra{\mathbf{0}}{\mathbf{0}} {U^{(r)}}^\dag ))]
  P(s_j|U_j^{(r)})\right],
\end{aligned}
\end{equation}
where $\mathbb{E}$ denotes the average over all local unitaries $U_k^{(r)}$'s and again $P(s_j|U_j^{(r)})=|\bra{s_j} U^{(r)}_j\ket{0}|^2$.
The latter corresponds to the $C_j$ introduced in Eq.~\eqref{eq:Cj} if $\Lambda =\bigotimes_{j=1}^{N} \Lambda_j$ and can be estimated from the calibration data as explained in Sec.~\ref{SM:calibration}, according to Eq.~\eqref{eq:estimatorCjSM}.
If we perform an average over all local random unitaries $U_k^{(r)}$  with $k\neq j$ (denoted  as $\mathbb{E}_{\{k \ne j\}}$), we can exploit the twirling identity for a single-qubit operator $O_j$, $\Phi^{(1)}_j(O_j)=\mathbb{E}[U_j^{(r)} O_j {U_j^{(r)}}^{\dag}]=\frac{\mathbb{1}}{2}\Tr(O_j)$, such that
\begin{equation}
\begin{aligned}\label{SM:phi1id}
\mathbb{E}_{\{k\neq j\}}
[U^{(r)}\ketbra{\mathbf{0}}{\mathbf{0}} {U^{(r)}}^\dag]
&=\mathbb{E}\left[U^{(r)}_1\ketbra{0}{0}{U^{(r)}_1}^\dagger \right]\otimes \cdots   \otimes U^{(r)}_j\ketbra{0}{0}{U^{(r)}_j}^\dag  \otimes \cdots \otimes \mathbb{E}\left[U^{(r)}_N\ketbra{0}{0}{U^{(r)}_N}^\dagger \right]\\
&=\Phi^{(1)}_1\left(\ketbra{0}{0} \right)\otimes \cdots \otimes U^{(r)}_j\ketbra{0}{0}{U^{(r)}_j}^\dag  \otimes \cdots \otimes \Phi^{(1)}_N\left(\ketbra{0}{0}\right)\\
&=
\mathbb{1}/2 \otimes \dots \otimes \mathbb{1}/2 \otimes U^{(r)}_j\ketbra{0}{0}{U^{(r)}_j}^\dag \otimes \mathbb{1}/2 \otimes\dots \otimes\mathbb{1}/2.
\end{aligned}
\end{equation}
and  write
\begin{equation}\label{eq:Cjmarginal}
\tilde C_j=\mathbb{E}\left[
\sum_{s_j = 0,1}\bra{s_j}
\tilde \Lambda_j(U^{(r)}_j\ketbra{0}{0} {U^{(r)}_j}^\dag ) \ket{s_j}
P(s_j|U_j^{(r)})\right],
\end{equation}
where  we have defined the `marginal channel' $\tilde \Lambda_j(\rho_j) = \Tr_{k \neq j} (\Lambda(\mathbb{1}/2\otimes \dots\otimes \mathbb{1}/2 \otimes \rho_j \otimes \mathbb{1}/2\otimes \dots\otimes \mathbb{1}/2))$.  Note that if $\Lambda=\bigotimes_{j=1}^{N} \Lambda_j$, we obtain $\tilde \Lambda_j=\Lambda_j$. 

Employing the same reasoning as in Eq.~\eqref{eq:twirlC}, we can average over the unitaries and use known results about two-copy twirling channels to find an expression for $\tilde{C}_j$:
\begin{equation}\label{SM:Gitilde}
\tilde{C}_j =\sum_{s_j = 0,1}
\bra{0}^{\otimes 2}
\Phi_j^{(2)} \left(\tilde \Lambda_j^*(\ketbra{s_j}{s_j})\otimes \ketbra{s_j}{s_j}\right)
\ket{0}^{\otimes 2}=
\frac{1}{6}
\sum_{s_j = 0,1}
(\bra{s_j} \tilde \Lambda^*_{j}(\ketbra{s_j}{s_j}) \ket{s_j}+\Tr[\Lambda^*_{j}(\ketbra{s_j}{s_j})])
  =
 \frac{1+\tilde{G}_j}{3}.
\end{equation}
Here $\tilde{G}_j=\frac{1}{2}\sum_{s_j}\bra{s_j} \tilde{\Lambda}_{j}(\ketbra{s_j}{s_j}) \ket{s_j}$ contains the information of the single qubit noise in term of a marginal channel, i.e. without the assumption of locality of the noise, and coincides with the one in Eq.~\eqref{SM:Gj}  in the case \mbox{$\Lambda=\bigotimes_{j=1}^{N} \Lambda_j$}. 

Let us proceed in a similar way for each pair of qubits $(j,j')$ of an $N$-qubit system in order to derive a quantity that also contains information about cross-talk errors.
In analogy with Eqs.~\eqref{eq:tilde_Cj} and~\eqref{eq:Cjmarginal}, for two qubits we define
\begin{eqnarray}
  \tilde{D}_{j,j'} &=&
\mathbb{E}\left[\sum_{s_j,s_{j'}}  
  \Tr [ \bra{s_{j},s_{j'}} \Lambda(U^{(r)} \ketbra{\0}{\0} {U^{(r)}}^\dag ) \ket{s_j,s_{j'}} ]
    P(s_j|U_j^{(r)})P(s_{j'}|U_{j'}^{(r)})\right]
   \nonumber \\
&=&
 \mathbb{E}\left[ \sum_{s_{j},s_{j'}}  
  \bra{s_{j},s_{j'}} \tilde{\Lambda}_{j,j'}\left({U^{(r)}_{j}} \otimes{U^{(r)}_{j'}}\ketbra{00}{00} {U^{(r)}_{j}}^\dag \otimes{U^{(r)}_{j'}}^\dag \right) \ket{s_{j},s_{j'}}
    P(s_j|U_j^{(r)})P(s_{j'}|U_{j'}^{(r)})\right] \nonumber \\
    &=&\mathbb{E}\left[
\sum_{s_{j},s_{j'}}
\bra{0}^{\otimes 4}
{{U^{(r)}_{j}}^{\dagger}}^{\otimes 2} {{U^{(r)}_{j'}}^{\dagger}}^{\otimes 2}
\left(\tilde{\Lambda}_{j,j'}^*(\ketbra{s_{j},s_{j'}}{s_{j},s_{j'}})\otimes \ketbra{s_{j},s_{j'}}{s_{j},s_{j'}}\right)
{U^{(r)}_{j}}^{\otimes 2} {U^{(r)}_{j'}}^{\otimes 2}
\ket{0}^{\otimes 4}\right],
\end{eqnarray}
where we have made use of the definition of the `marginal channel' $\tilde \Lambda_{j,j'}(\rho_j \otimes\rho_{j'})$ defined as
\begin{equation}
   \tilde \Lambda_{j,j'}(\rho_j\otimes\rho_{j'}) = \mathrm{Tr}_{k\neq j, j'}\left(\Lambda( \mathbb{1}/2 \otimes \dots \otimes \mathbb{1}/2 \otimes \rho_{j} \otimes \mathbb{1}/2 \dots \mathbb{1}/2 \otimes \rho_{j'} \otimes \mathbb{1}/2\dots\otimes \mathbb{1}/2) \right)
\end{equation}
that exploits the same reasoning as Eq.~\eqref{SM:phi1id}. This quantity can be estimated from the calibration data $\tilde{C}_j$ by extending the estimators in Eq.s~\eqref{eq:estimatorCjSM}-\eqref{eq:CCRestimator} to two-qubit measurements.

As previously done for the single qubit quantity $\tilde{C}_j$ we can now perform explicitly the average over the unitaries on the pair of qubits $(j,j')$ exploiting the appropriate twirling channel identities.
In particular, we can write 
\begin{equation}
\tilde{D}_{j,j'}=\ \sum_{s_{j},s_{j'} = 0, 1}
\bra{0}^{\otimes 4}
\Phi_{j,j'}^{(2)}
\left(\tilde \Lambda_{j,j'}^*\left(\ketbra{s_{j},s_{j'}}{s_j,s_{j'}}\right)\otimes \ketbra{s_{j},s_{j'}}{s_j,s_{j'}}\right)
\ket{0}^{\otimes 4}=
\bra{0}^{\otimes 4}
\Phi_{j,j'}^{(2)}
\left(Q_{j,j'}\right)
\ket{0}^{\otimes 4},
\end{equation}
where we have defined $Q_{j,j'}=\sum_{s_j,s_{j'}}\tilde \Lambda_{j,j'}^*\left(\ketbra{s_j,s_{j'}}{s_j,s_{j'}}\right)\otimes \ketbra{s_j,s_{j'}}{s_j,s_{j'}}$. Here we also introduced $\Phi_{j,j'}^{(2)}$ such that $\Phi_{j,j'}^{(2)}(Q_j\otimes Q_{j'}) = \mathbb{E}\big[ {{U^{(r)}_{j}}^{\dagger}}^{\otimes 2} {{U^{(r)}_{j'}}^{\dagger}}^{\otimes 2} (Q_j\otimes Q_{j'}) {U^{(r)}_{j}}^{\otimes 2} {U^{(r)}_{j'}}^{\otimes 2} \big] = \Phi_j^{(2)}(Q_j)\otimes \Phi_j^{(2)}(Q_{j'})$, that can be extended linearly to non-product observables $Q_{j,j'}$. Using the twirling formula in Eq.~\eqref{eq:twirlingformula} and working out the analytics, one obtains (with implicit identity operators)
\begin{equation}
\tilde{D}_{j,j'}=
\bra{0}^{\otimes 4}
\Phi_{j,j'}^{(2)}
\left(Q_{j,j'}\right)
\ket{0}^{\otimes 4}
=
\frac{1}{36}
\left[
\Tr(Q_{j,j'})+
\Tr(\S_jQ_{j,j'})+
\Tr(\S_{j'}Q_{j,j'})+
\Tr(\S_j\S_{j'}Q_{j,j'})
\right].
\end{equation}
We can then compute
\begin{equation}
\begin{aligned}
\Tr(Q_{j,j'})&=4, \\
 \Tr(\S_j Q_{j,j'})&=
\sum_{s_j,s_{j'}}
\bra{s_j}
\Tr_{j'}[\tilde{\Lambda}_{j,j'}^*
(\ketbra{s_j,s_{j'}}{s_j,s_{j'}}])
\ket{s_j}
=
 2\sum_{s_j}\bra{s_j}\tilde{\Lambda}_j\left( \ketbra{s_j}{s_j} \right)\ket{s_j}=4\tilde G_j, \\
\Tr(\S_{j'} Q_{j,j'})&=4\tilde G_{j'}, \\
\Tr(\S_j\S_{j'} Q_{j,j'})&=\sum_{s_j,s_{j'}}\bra{s_js_{j'}}\tilde{\Lambda}_{j{j'}}\left(\ketbra{s_j,s_{j'}}{s_j,s_{j'}}\right)\ket{s_js_{j'}}\equiv 4\tilde G_{j,j'},
\end{aligned}
\end{equation}
where
      \begin{equation}\label{SM:Gijtilde}
      \tilde G_{j,j'}=
      \frac{1}{4}
      \sum_{s_j,s_{j'}} \bra{s_j,s_{j'}}\tilde \Lambda_{j,j'}\left(\ketbra{s_j,s_{j'}}{s_j,s_{j'}}\right)\ket{s_j,s_{j'}}.
      \end{equation}
Eventually, we arrive at the following expression for $\tilde{D}_{j,j'}$
\begin{align}    \tilde{D}_{j,j'}    =\frac{1}{9}
    \left(
    1+\tilde G_j+\tilde G_{j'}+\tilde G_{j,j'}
    \right).
\end{align}
      
Estimating $\tilde{C}_j$, $\tilde{D}_{j,j'}$, we have thus experimental access to the terms $\tilde G_j,\tilde G_{j,j'}$ that contain information about the noise channel $\Lambda$.
Both of them are equal to 1 in the absence of noise ($\tilde{G}_j=\tilde{G}_{j,j'}=1$) and if $\tilde G_{j,j'} \neq \tilde G_j \tilde{G}_{j'}$ then $\Lambda \neq \bigotimes_{j=1}^{N} \Lambda_j$, i.e. the error is not local.
Thus, we introduce the following quantity
\begin{equation}
    \tilde{R}=\tilde{G}_{j,j'}-\tilde{G}_j\tilde{G}_{j'}
\end{equation}
as a proxy of cross-talk effects. In particular, $\tilde{R}\neq 0$ witnesses the presence of cross-talk in the system according to the previous reasoning, namely $\tilde{R} \neq 0$ implies that $\Lambda$ is not factorized.  Let us remark here that $\tilde{R}= 0$ cannot exclude the presence of cross-talk. In fact, there exist noise channels $\Lambda$ that introduce cross-talk effects but satisfy the condition $\tilde{R}= 0$. 
In the following, we provide an example of a noise channel that could model measurement errors in simple cases and show that $\tilde{R}$ is able to detect cross-talk noise contributions in this case. Furthermore, employing this noise model, we observe that such contributions are negligible compared to the local ones.

\subsection{Application to a two-qubit readout error model}
Since the analysis presented in Fig.~\ref{fig6} suggests that the error in the platform is mostly due to readout, in this section we focus on a simple readout error model.
We consider the following noise channel $\Lambda_{j,j'}=\Lambda^{(2)}_{j,j'}\circ (\Lambda^{(1)}_j \otimes \Lambda^{(1)}_{j'})$ for two qubits $(j,{j'})$ where:
\begin{equation}\label{eq:errorchannel}
\begin{aligned}
    &\Lambda^{(2)}_{j,j'}(\rho_{j,j'})
    =
    (1-p_{\textrm{NL}})\rho_{j,j'}
    + p_{\textrm{NL}} \sigma_j^x\sigma_{j'}^x \rho_{j,j'}\sigma_{j}^x\sigma_{j'}^x;\\
    & \Lambda^{(1)}_{j}(\rho_{j})
    =
    (1-p^{(j)}_\textrm{L})\rho_{j}
    + p^{(j)}_\textrm{L} \sigma_j^x\rho_{j}\sigma_{j}^x.   
\end{aligned}
\end{equation}
This model contains cross-talk errors with probability $p_{\textrm{NL}}$ -- namely, correlated bit-flips (which could model measurement errors) for qubits $j$ and $j'$ -- and single qubit bit-flips with probability $p^{(k)}_\textrm{L}$, which could a priori be different for each qubit $k=j,j'$. 
In the low noise limit $p_{\textrm{NL}},p^{(k)}_{\textrm{L}}\ll 1$, at first order, one can write the noise channel $\Lambda_{j,j'}$ as
\begin{equation}
    \Lambda_{j,j'}(\rho_{j,j'})
    \simeq
    (1-p_{\textrm{NL}}-p^{(j)}_\textrm{L}-p^{(j')}_\textrm{L})\rho_{j,j'}
     + p_{\textrm{NL}} \,\sigma_j^x\sigma_{j'}^x \rho_{j,j'}\sigma_{j}^x\sigma_{j'}^x
     + p^{(j)}_\textrm{L} \sigma_j^x\rho_{j,j'}\sigma_{j}^x
     + p^{(j')}_\textrm{L} \sigma_{j'}^x\rho_{j.j'}\sigma_{j'}^x.
\end{equation}
Employing the definitions of Sec.~\ref{sec:derivationlocality} one gets
\begin{align}\label{eq:Gasp1}
&\tilde{G}_{j,j'}
\simeq1-p_{\textrm{NL}}-p^{(j)}_\textrm{L}-p^{(j')}_\textrm{L},\\\label{eq:Gasp2}
&\tilde{G}_j
\simeq1-p_{\textrm{NL}}-p^{(j)}_\textrm{L},\\\label{eq:Gasp3}
&\tilde{G}_{j'}
\simeq1-p_{\textrm{NL}}-p^{(j')}_\textrm{L},
 \end{align}
and that gives
\begin{align}\label{eq:tildeR_model}
\tilde R
\simeq
p_{\textrm{NL}}.
 \end{align}

For any small values of $p_L$, $\tilde R$ is uniquely related to the cross-talk probability $p_{\text{NL}}$. Furthermore, $\tilde{R}\neq 0$ when the nonlocal term $p_{\textrm{NL}}$ is different from zero and can be used to detect the cross-talk noise according to noise model employed.
Hence, in the following we will adopt this noise model to investigate the strength of the cross-talk error in the quantum platform we have used in this work.
 
\begin{figure}
    \centering
    \includegraphics[width=0.5\linewidth]{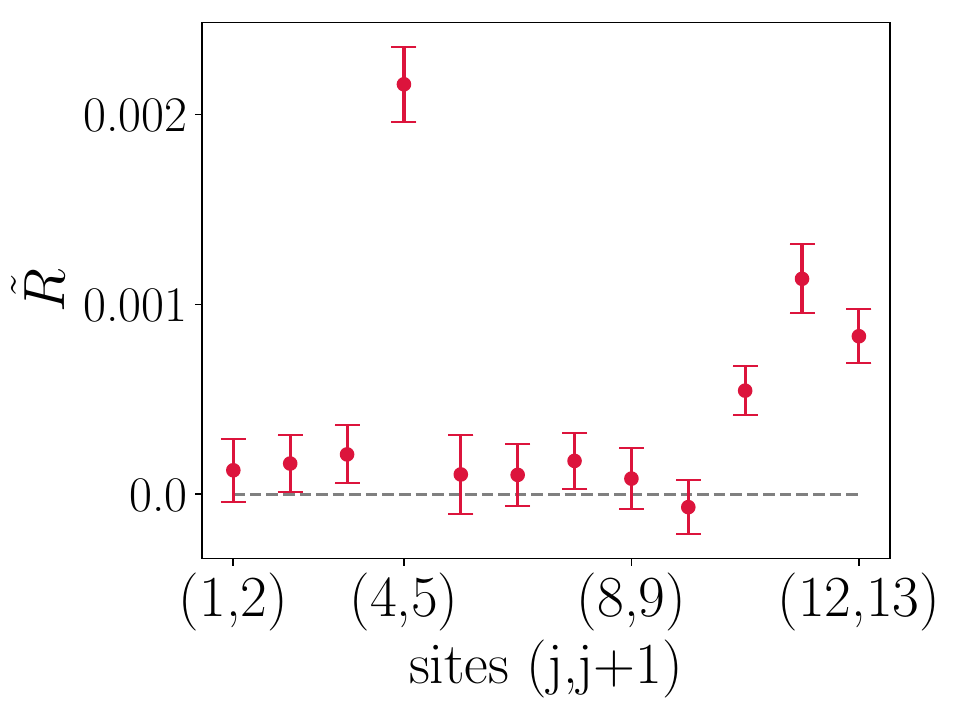}
    \caption{
    We estimate $\tilde{R}$  for neighbouring qubits $(j,j')$ as $\hat{\tilde{R}}= \frac{1}{N_I}\sum_{i=1}^{N_I} \hat{\tilde{R}}^{(i)}$, where $\hat{\tilde{R}}^{(i)}=\hat{\mathcal{G}}^{(i)}_{j,j'}-\hat{\mathcal{G}}^{(i)}_{j}\hat{\mathcal{G}}^{(i)}_{j'}$. The error bars are estimated as the standard deviation of the mean. A value not compatible within errorbars with 0 (horizontal grey line) means that cross-talk between the qubits is present. The measurement budget is the one described in Sec.~\ref{sec:budget} for the calibration experiment.}
    \label{fig7}
\end{figure}
\subsection{Experimental investigation on the employed platform}\label{localityofnoiseexperiment}
Here, we study the locality of the noise on the platform `ibm\_prague' that we have used to prepare the states of interest and measure the QFI. We employ the quantity $\tilde{R}$ and estimate it using the calibration data for the 13-qubit state collected according to the indications in Sec.~\ref{sec:EP}.
In Fig.~\ref{fig7} we show $\tilde{R}$ for neighbouring qubits, averaged over the iterations $i=1,\dots,N_I$, namely $\tilde{R}=\frac{1}{N_I}\sum_{i=1}^{N_I} \tilde{R}^{(i)}$. The error bars are estimated as the standard deviation of the mean of the different estimates.
A value not compatible with 0 (horizontal grey line) witnesses the presence of cross-talk, namely $\tilde{R} \neq0 \Rightarrow \Lambda \neq \bigotimes_{i=1}^N\Lambda_i$. We observe that it is the case for the pairs of qubits $(4,5)$, $(10,11)$, $(11,12)$, $(12,13)$. 

To estimate the strength of the cross-talk with respect to the local noise in the system we employ the noise model introduced in the previous section, Eq.~\eqref{eq:errorchannel} to compute $p_{\textrm{L}}^{(k)}$ and $p_{\textrm{NL}}$ from the measured values of $\tilde{G}_{j,j'}$ and $\tilde{G}_{j}$.
At first order in $p_{\textrm{NL}},p^{(k)}_{\textrm{L}}$ -- in the limit $p_{\textrm{NL}},p^{(k)}_{\textrm{L}}\ll 1$ -- one obtains
\begin{align}\label{eq:pformulas1}
        &p^{(j)}_\text{L}\simeq\tilde{G}_{j'}-\tilde{G}_{j,j'},\\\label{eq:pformulas2}
        &p^{(j')}_\text{L}\simeq\tilde{G}_{j}-\tilde{G}_{j,j'},\\\label{eq:pformulas3}
        &p_\text{NL}\simeq \tilde{R}.
\end{align}
by inverting Eqs.~\eqref{eq:Gasp1}--\eqref{eq:tildeR_model}. 
\begin{table}[]
    \centering
    \def\arraystretch{1.5}%
    \begin{tabular}{|C{1.7cm}|C{1.7cm}|C{1.7cm}|C{1.7cm}|C{1.7cm}|C{1.7cm}|C{1.7cm}|C{1.7cm}|C{1.7cm}|}
    \hline
            pair (j,j') & $\tilde{G}_j$  & $\tilde{G}_{j'}$ & $\tilde{G}_{j,j'}$  & $p^{(j)}_\text{L}$ & $p^{(j')}_{\text{L}}$ &  $p_{\text{NL}}\simeq\tilde{R}$\\
            \hline
 $(1,2)$ & 0.9775(3) & 0.9783(2)  & 0.9565(4) &  0.0218(5) & 0.0210(3) & 0.0001(1)\\
\hline
$(2,3)$ & 0.9783(2)  & 0.9873(1) & 0.9661(3)  & 0.0212(4) & 0.0122(3)& 0.0002(1)\\
\hline
$(3,4)$ & 0.9873(1) & 0.9756(2) &  0.9634(3) &  0.0121(4) & 0.0238(4)  & 0.0002(1)\\
\hline
$(4,5)$ & 0.9756(2)  & 0.9661(4) &  0.9447(5)  & 0.0214(5) & 0.0308(6) & 0.0022(1)\\
\hline
$(5,6)$ & 0.9661(4) & 0.9847(10) &  0.9515(10)  & 0.0332(11) & 0.0146(14) & 0.0001(1)\\
\hline
$(6,7)$ & 0.9847(10)  & 0.9754(2) &  0.9606(10) & 0.0148(14) & 0.0240(10) & 0.0001(1)\\
\hline
$(7,8)$ & 0.9754(2)  & 0.9843(2) &  0.9604(2) & 0.0239(3) & 0.0150(3) & 0.0002(1)\\
\hline
$(8,9)$ & 0.9843(2) & 0.9815(2)  & 0.9662(3)  & 0.0152(3) & 0.0181(3) & 0.0002(1)\\
\hline
$(9,10)$ & 0.9815(2) & 0.9843(2)  & 0.9669(3)  &  0.0182(4)  & 0.0154(3) & -0.0001(1)\\
\hline
$(10,11)$ & 0.9844(2)  & 0.9863(2)  & 0.9713(3)  &0.0149(4) & 0.0129(5) & 0.0005(1)\\
\hline
$(11,12)$ & 0.9863(2)  & 0.9727(3)  & 0.9595(3)  &  0.0121(4) & 0.0267(5) & 0.0030(1)\\
\hline
$(12,13)$ & 0.9717(3) & 0.9892(1)  & 0.9621(3) & 0.0271(4) & 0.0096(4) & 0.0008(1)\\
\hline
    \end{tabular}
    \caption{Table containing the experimentally measured values of $\tilde{G}_j$, $\tilde{G}_{j'}$, $\tilde{G}_{j,j'}$ and $p_{\text{NL}}$, $p^{(k)}_\text{L}$ ($k=j,j'$) calculated according to Eqs.~\eqref{eq:Gasp1}--\eqref{eq:tildeR_model} and using the estimators $\hat{\mathcal{G}}_{j}$, $\hat{\mathcal{G}}_{j,j'}$. The number in parentheses is the numerical value of the statistical error referred to the corresponding last digits of the result.}
    \label{tab:table}
\end{table}
Plugging in these equations the experimental values of $\tilde{G}_{j,j'}$ and $\tilde{G}_{j}$ it is possible to compute the probability ratio $p_{\text{NL}}/p^{(k)}_{\text{L}}$ that is informative of the relative strength of nonlocal noise. 
Such as for $\tilde{R}$, the measured values of $\tilde{G}_{j,j'}$ and $\tilde{G}_{j}$ are an average over the estimates of the different iterations $i=1,\dots,N_I$ and their error bars are calculated as the standard deviation of the mean. We employ the estimators $\hat{\mathcal{G}}_j$ and $\hat{\mathcal{G}}_{j,j'}$ discussed in the main text and Sec.~\ref{SM:Gijtilde}.

We give the experimental results for any pair of neighbouring qubits in Tab.~\ref{tab:table}.
We observe that in the illustrative case of qubits $(4,5)$ -- where $\tilde{R}=0.002$ signals the presence of nonlocal noise -- we obtain $p_{\text{NL}}/p^{(k)}_\text{L}\simeq 10^{-1}$ for both $k=j,j'$. More in general, for pairs $(4,5)$, $(11,12)$ and $(12,13)$, $p_{\text{NL}}$ is not compatible with zero within errors. However, given that $p_{\text{NL}}/p^{(k)}_\text{L}\lesssim 10^{-1}$ we can conclude that the amount of cross-talk error in our platform would not harm the robust shadow protocol that we employ, as investigated numerically in Ref.~\cite{Chen2020}.
The dominant source of error, under the assumptions of our noise model, corresponds to local measurement errors, which can be corrected faithfully via local robust shadows.

\section{Further experimental results}\label{moreexpresults}
In this appendix we study the estimator in Ref.~\cite{Yu2021} in comparison with our bounds $F_n$. Then, we provide more experimental results on the QAOA protocol and a comparison of two different methods of calibration, namely calibrating the circuit only once at the beginning of the experiment or repeating the calibration in each iteration to follow the time fluctuations.

\subsection{Comparison with previous work in Ref.~\cite{Yu2021}}\label{SM:DGvsQFI}
In this section, we post-process our recorded robust RM measurement data to obtain another lower bound to the QFI that has been previously estimated using standard RM formalism.
The lower bound of interest is defined in Refs.~\cite{Cerezo2021, Yu2021} as a function of the quantum states $\rho_{\theta}$ and $\rho_{\theta+d\theta}$, as
 \begin{equation}
    \mathcal{F}_G(\rho_{d\theta }) \equiv \frac{D_G(\rho_{\theta}, \rho_{\theta +d\theta})}{d\theta ^2} 
    = \frac{8\left[1- \Tr(\rho_{\theta} \rho_{\theta+d\theta}) + \sqrt{(1-\Tr(\rho_{\theta}^2))(1-\Tr(\rho_{\theta +d\theta}^2))}\right]}{d\theta^2}. \label{eq:Fg}
 \end{equation}
It is important to note for the above lower bound that $\lim_{d\theta \to 0} \mathcal{F}_G(\rho_{d\theta}) = F_0$ \cite{Cerezo2021} and estimating it requires one to be able to distinguish between a state $\rho_{\theta}$ and its neighbor $\rho_{\theta + d\theta}$ that encode an unknown parameter $\theta$.

In our analysis, we estimate unbiased estimators for each of the terms in RHS of Eq.~\eqref{eq:Fg} according to U-statistics, as detailed in the main text (Sec.~\ref{sec:datacompression}). 
The $\theta$ parametrized state is defined as $\rho_{\theta} = e^{-i\theta A} \rho e^{i\theta A}$ with $A = \frac12 \sum_j \sigma_z^{(j)}$. 
We consider here $\theta = 0$ for simplicity as done in Ref.~\cite{Yu2021}.
Additionally, compared to Ref.~\cite{Yu2021} that prepared experimentally the two parametrized states $\rho_{\theta}$ and $\rho_{\theta +d\theta}$, with the robust classical shadow formalism, we can estimate $D_G(\rho_{\theta}, \rho_{\theta+d\theta})$ by performing this step classically during the post-processing stage. 
We remark that robust classical shadows defined in Eq.~\eqref{eq:robshadows} satisfy $\rho_\theta = \expp[e^{-i\theta A} \tilde{\rho}^{(r_i)} e^{i\theta A}]$ with the average taken over the applied random unitaries and measurements.

\begin{figure}
    \centering
    \includegraphics[width=0.9\linewidth]{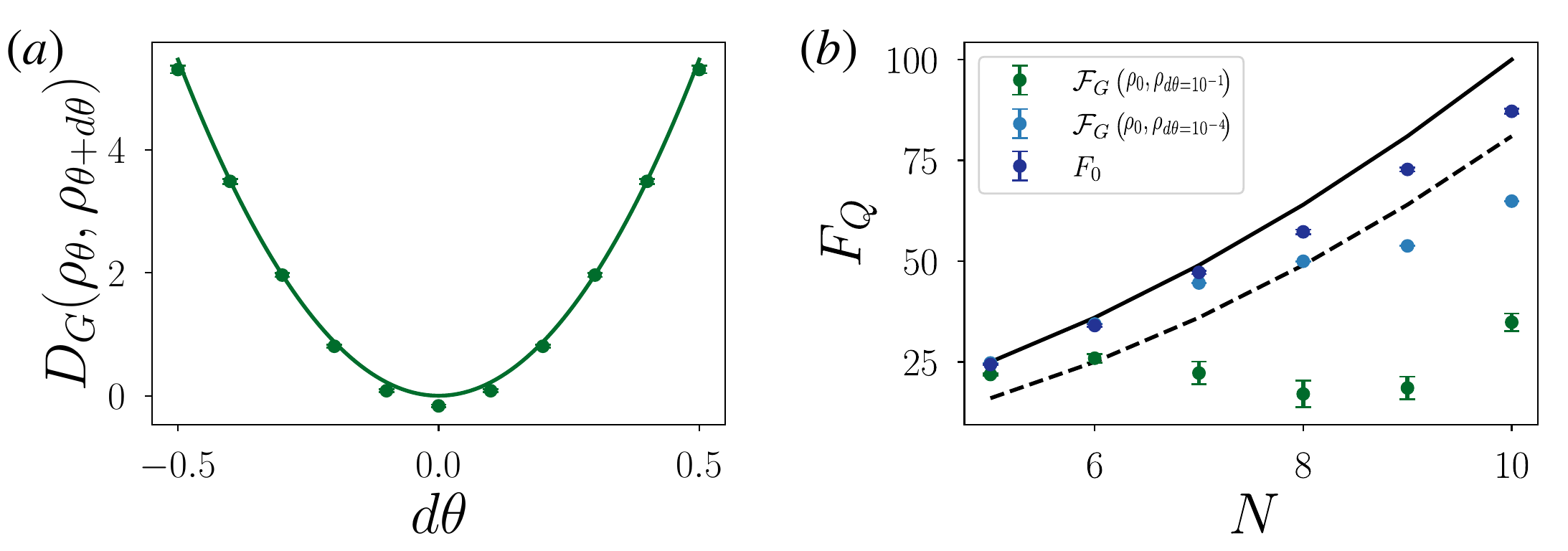}
    \caption{$(a)$ Modified Bures distance $D_G$ (and quadratic fit) for a GHZ state comprised of $N=5$ qubits. $(b)$ Comparison between $\mathcal{F}_G$ (for different values of $d\theta$) and the lower bound $F_0$ as a function of the number of qubits $N$. The solid black line corresponds to $F_Q=N^2$, the dashed black line to the threshold above which the state can be considered GME.}
    \label{fig:DG}
\end{figure}

Fig.~\ref{fig:DG} summarizes the analysis for our experimental data. We consider the RM data taken after preparing a $N-$qubit GHZ states. 
Firstly, as shown in Fig.~\ref{fig:DG}$(a)$, we estimate the modified Bures distance $D_G(\rho_{\theta_1}, \rho_{\theta_2})$ as a function of $d\theta = \theta_2 - \theta_1$ for a 5-qubit GHZ state~\cite[Fig.~2(b) for 4-qubit GHZ state]{Yu2021}. 
We then perform a polynomial fit as shown by the solid line in Fig.~\ref{fig:DG}$(a)$. The coefficient of the quadratic term of this fit provides an estimation of $\mathcal{F}_G(\rho_{d\theta})$.
This is extracted for qubit sizes ranging from 5 to 10 in Fig.~\ref{fig:DG}$(b)$ for two different values of $d\theta$. 
We observe, that $\mathcal{F}_G(\rho_{d\theta = 10^{-1}}) \leq \mathcal{F}_G(\rho_{d\theta = 10^{-4}}) \leq F_0$ for all qubit sizes. 
It is important to note that in order to obtain values of $\mathcal{F}_G$ that are comparable to $F_0$, one has to perform the procedure employed in Ref.~\cite{Yu2021} by encoding very small parameter shifts that become extremely challenging experimentally.

\subsection{Ground state of the TFIM at the critical point}\label{SM:QAOA}
\begin{figure}
    \centering
    \includegraphics[width=\linewidth]{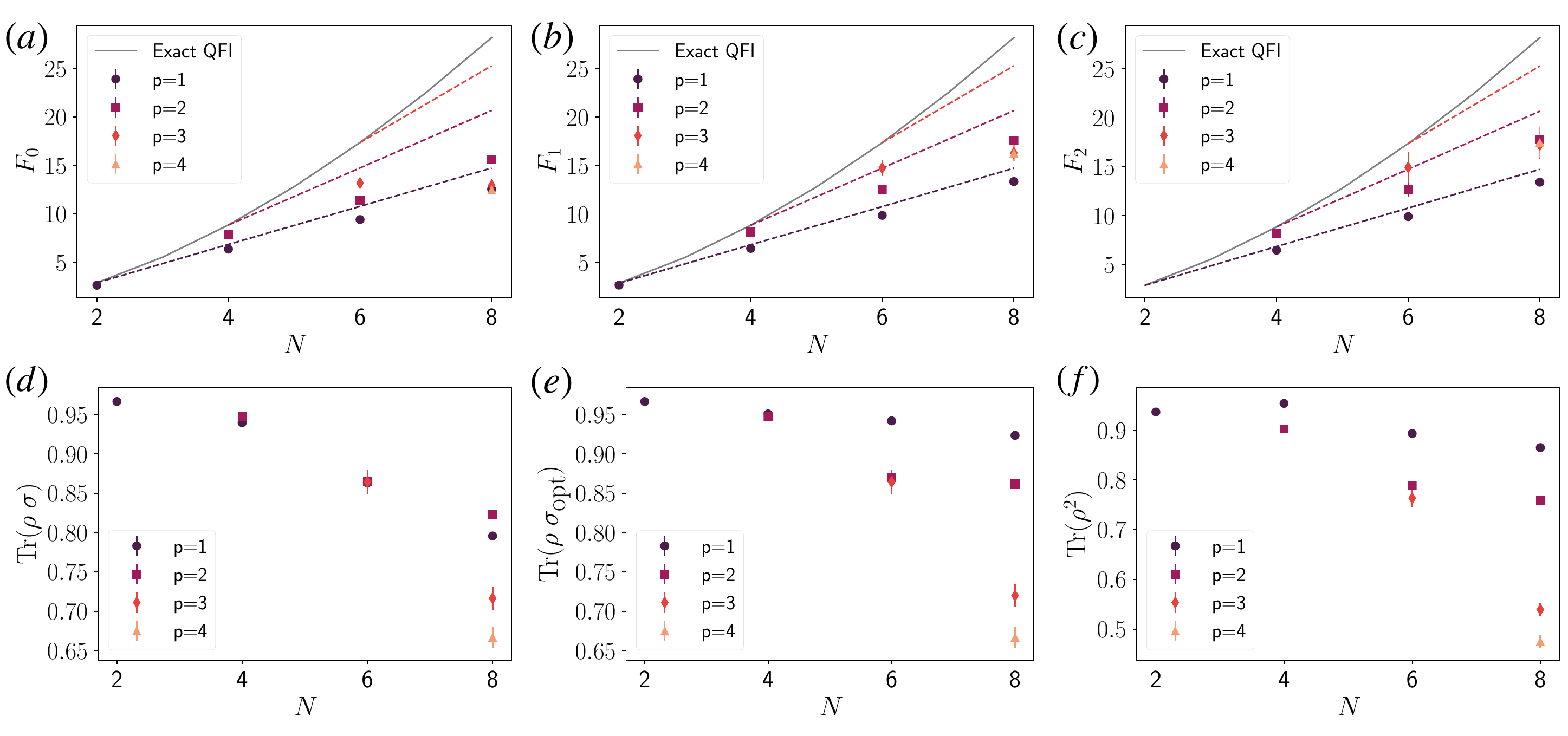}
    \caption{More experimental results on the TFIM at the critical point. In the upper panels we present results for $(a)$ $F_0$, $(b)$ $F_1$ and $(c)$ $F_2$. In the lower panels  we show $(d)$ the fidelity with respect to the real ground state $\sigma$, $(e)$ the fidelity with respect to the state $\sigma_\textup{opt}$ obtained by classical optimization and experimentally prepared through QAOA at fixed depth $p$, and $(f)$ the purity $\Tr(\rho^2)$ of the prepared state. The results for $F_1$ have already been presented in the main text. The measurement protocol details are described in Sec.~\ref{sec:EP}. In all the plots, $p$ is the number of layers in the circuit, i.e. the depth of the circuit. The solid black lines correspond to the exact value of the QFI $F_Q$. The colored dashed line denote the theoretical value of the QFI at fixed depth.}
    \label{fig9}
\end{figure}
In this section we give additional experimental results concerning the ground state of the TFIM at $h=1$, prepared with the variational circuit described in the main text.
We have showed explicitly the value of the bound $F_1$ in the main text.
In the upper plots of Fig.~\ref{fig9} we show $F_0$, $F_1$ and $F_2$.
The lines in $(a)$, $(b)$ and $(c)$ denote the theoretical QFI for a fixed circuit depth $p$.
In the lower plots we show the fidelities of state preparation and purity in the lower ones.
In particular, in $(d)$ we show the fidelity with respect to the real ground state of the TFIM at the critical point, computed with exact diagonalization. We observe that increasing the depth $p$ the fidelity generally drops, because the more layers the higher the noise in the system. The only exception is the case of depth $p=2$ for $N=8$, the rationale behind it is that the approximation of the ground state for $p=1$ ad $N=8$ is very poor, and thus, even in the presence of noise, for $p=2$ the prepared state is closer to the true one. 
The worsening of state preparation with $p$ is also evident in $(e)$ where we plot the fidelity with respect to the state prepared in the case of a noiseless QAOA. The fidelity is always better for $p=1$, but we remind the reader that the prepared state is not faithful to the ground state of the TFIM for larger system sizes.
Finally in $(f)$ we show the purity of the prepared states. Again we observe that it drops with increasing $p$ due to the presence of noise. The final state should be a pure state in the ideal scenario, i.e. $\Tr (\rho^2)=1$. Here we observe that increasing the number of layers tends to decrease the purity of the prepared state, e.g. for $p=4$ and $N=8$ one has $\Tr (\rho^2) \sim 0.5$.

\subsection{Comparison of error mitigation protocols}\label{SM:batchvsfull}
\begin{figure}
    \centering
    \includegraphics[width=0.8\linewidth]{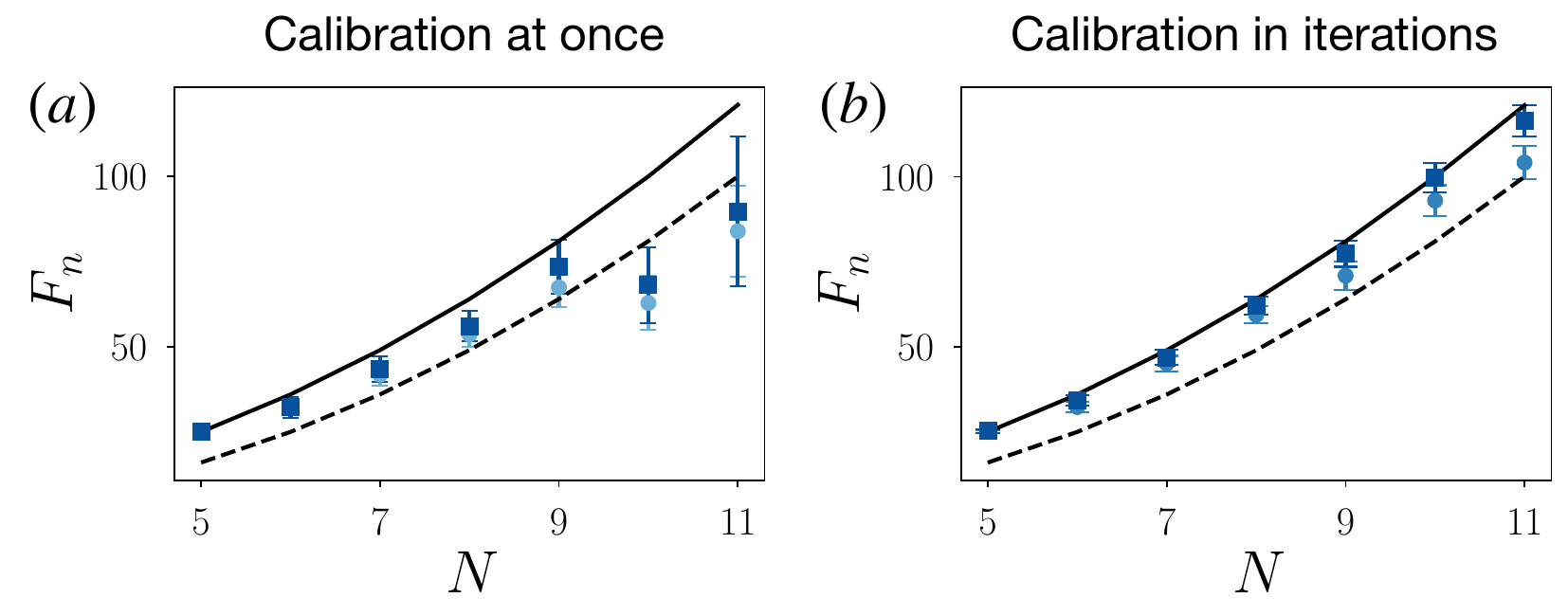}
    \caption{Comparison of of different methods of calibration. $F_0$ and $F_1$ (light to dark) when $(a)$ the calibration of the device is performed at the beginning of the whole experimental run, $(b)$ the calibration is performed in each iteration according to the experimental protocol described in the main text. The measurement budget of both experiments is the same and is detailed in Sec.~\ref{sec:budget}. The solid line is the exact value of the QFI ($F_Q = N^2$) for pure GHZ states. The dashed black line corresponds to the entanglement witness \mbox{$\Gamma (N,k=N-1) = (N-1)^2$} above which the state can be claimed to be GME.}
    \label{fig10}
\end{figure}
In this section, we show experimental evidence that calibrating in iterations is more efficient than calibrating it once at the beginning of the experiment. In Fig.~\ref{fig10} we compare the estimation of the bound of the QFI when the calibration of the device is performed at the beginning of the whole experimental procedure or according to our prescriptions. We present the error mitigated experimental estimations of $F_0$ and $F_1$ (light to dark).
In Fig.~\ref{fig10}$(a)$ the calibration is performed at the beginning. We observe that the robust estimation for larger system sizes is not compatible with the $\sim N^2$ scaling predicted by the theory, and that it does not violate the witness of  $(N-1)^2$ that validates GME.
In Fig.~\ref{fig10}$(b)$ we present the same experimental results of the main text for $F_0$ and $F_1$. As already commented, we observe the $\sim N^2$ scaling of QFI and witness GME.
The discrepancy is due to the fluctuating gate and readout errors in the quantum processors that affect the reliability of the results when the experimental run takes long times, that is for larger $N$.

\section{Numerical investigations}\label{SM:numerics}
In this section we study the behavior of the classical Fisher information in comparison with the quantum Fisher information. We investigate the scaling of $\hat{\mathcal{G}}_j$ as a function of the number of unitaries $N_U$ for different values of the readout error and the scaling of the required number of measurements to achieve a given level of statistical errors on our highest measured bound $F_2$.
Lastly, we present classical numerical experiment for GHZ states prepared without any state preparation errors.

\subsection{Classical and Quantum Fisher information comparison in noisy GHZ states}\label{SM:CFIvsQFI}
The Fisher information is a fundamental concept in statistics and information theory~\cite{fisher1922mathematical,fisher1925theory}. It measures the amount of information that a random variable carries about an unknown parameter when sampled from a given probability distribution and plays a crucial role in the field of metrology.
In particular, in the context of
estimating an unknown parameter $\theta$ encoded in a quantum state $\rho$, it has been used to show that the precision of the
measurement could go beyond the shot-noise limit~\cite{Pezze2018}. 

The choice of the measurement setting significantly influences the accuracy of the estimation process. Optimal choices are characterized by measurement results exhibiting a statistical distribution that is highly sensitive to variations in $\theta$.
Indicating as $P(\mu |\theta)$ the probability of a measurement result $\mu$ given that the parameter has the value $\theta$, the classical Fisher information can be written as:
\begin{equation}
    F(\theta)=\sum_\mu \frac{1}{P(\mu|\theta)} \left( \frac{\partial P(\mu|\theta)}{\partial \theta}\right)^2.
\end{equation}
An upper bound to the Fisher information is obtained by maximizing the previous equation over all possible generalized measurements settings~\cite{Braunstein1994} and corresponds to the quantum Fisher information (QFI), defined in Eq.~\eqref{eq:qfi} of the main text.

Therefore, by definition, the classical Fisher information (CFI) is upper-bounded by the QFI. However, it has been readily employed in quantum experiments to prove the presence of multipartite entanglement or enhanced metrological sensitivity with respect to the classical cases~\cite{Strobel2014, Bohnet2016,Pezze_2016_PNAS}.
The latter is due to the fact that it is easier to measure in general, since it does not need the full spectral resolution of the density matrix, even though one is faced with the challenge of finding good measurement observables and measurement basis.
In this work we proposed a method to estimate directly the QFI, that is intrinsically not affected by the choice of the measurement basis, employing a converging series of lower bounds to it.
In this section, for the sake of completeness we draw a comparison between the CFI, for some fixed choice of measurement, and the QFI in noisy GHZ states.

The CFI, in this instance, can be calculated analytically. We write the GHZ state of a $N$-qubits system as $\ket{\psi}=(\ket{0}^{\otimes N}+\ket{1}^{\otimes N})/\sqrt{2}$. We perform the evolution under the operator $U=e^{-i \theta A}$ with $A=\frac{1}{2}\sum_i \sigma_j^{z}$ and we measure the qubits along the y-axis. We remark that the latter means applying a  phase gate $S^{\dagger}=\left(\begin{smallmatrix}
    1 & 0 \\
    0 & i
\end{smallmatrix}\right)^\dagger$ and a Hadamard gate $H=\frac{1}{\sqrt{2}}\left(\begin{smallmatrix}
    1 & 1 \\
    1 & -1
\end{smallmatrix}\right)$ to the state, before measuring along the z-axis.
The probability of obtaining as an outcome an $N$-bit string $\s$ is given by
\begin{align}
    P(\s|\theta)&=|\bra{\s}(H S^\dag)^{\otimes N} e^{-i\theta A}\ket{\psi}|^2.
\end{align}
Performing the calculation one obtains
\begin{equation}
    (H S^\dag)^{\otimes N} e^{-i\theta A}\ket{\psi}
    =\frac{1}{\sqrt{2}}(H S^\dag)^{\otimes N}
    \left(
    e^{-i\theta N/2}
    \ket{0}^{\otimes N}
    +
    e^{i\theta N/2}
    \ket{1}^{\otimes N}
    \right)
    =
    \frac{1}{\sqrt{2^{N+1}}}
    \sum_{\s}
    \left(
    e^{-i\theta N/2}
    +
    e^{i\theta N/2}
    e^{i\pi N/2}
    e^{i\pi |\s|}
    \right)    \ket{\s}
\end{equation} 
leading to 
\begin{equation}
\begin{aligned}
    P(\s|\theta)&=
    \frac{1}{2^{N}}
    \left[
    1+(-1)^{|\s|}\left(\cos(\theta N)\cos\left(\frac{\pi}{2}N\right)+\sin(\theta N)\sin\left(\frac{\pi}{2}N\right)\right)\right].
\end{aligned}
\end{equation}
We observe that in $\theta=\frac{\pi}{2N}\equiv \theta_0$ and assuming that $N$ is even on gets
\begin{equation}
    \begin{aligned}
        P(\s|\theta_0)&=
        \frac{1}{2^{N}},\\
        \frac{\partial P(\s|\theta)}{\partial \theta}\Bigg|_{\theta=\theta_0}&=
        -\frac{N}{2^{N}}
        (-1)^{|\s|+N/2}.
    \end{aligned}
\end{equation}
Eventually, the CFI reads
\begin{equation}
F(\theta_0)=\sum_\s \frac{1}{P(\s|\theta_0)} \left( \frac{\partial P(\s|\theta)}{\partial \theta}\Bigg|_{\theta=\theta_0}\right)^2=N^2.
\end{equation}
Therefore, employing this measurement scheme, the CFI reaches the maximum possible value for a system comprised of $N$ qubits and coincides with the QFI.
We will show here that in the presence of global depolarization, the CFI (obtained with the same, fixed, measurement as considered above) decreases faster than the QFI, whose explicit functional dependence with respect to the strength of the noise has been given in Eq.~\eqref{eq:qfidepoMT} and discussed in Sec.~\ref{sec:depoQFI}.
In the presence of global depolarization of strength $p_D$, then $P(\s|\theta)$ calculated above is changed to $(1-p_D)P(\s|\theta) + p_D \frac{1}{2^{N}}$, so that now one has
\begin{equation}
    \begin{aligned}
        P(\s|\theta_0)&=
        \frac{1}{2^{N}}\\
        \frac{\partial P(\s|\theta)}{\partial \theta}\Bigg|_{\theta=\theta_0}&=-(-1)^{|\s|+N/2}(1-p_D)\frac{N}{2^{N}}
    \end{aligned}
\end{equation}
which leads to the following value for the CFI 
\begin{equation}\label{eq:CFIdepo}
    F(\theta_0) =
    \frac{1}{2^N}
    \sum_s (1-p_D)^2 N^2=(1-p_D)^2 N^2
\end{equation}
With respect to Eq.~\eqref{eq:qfidepoMT}  we observe that the CFI decays faster (as $\sim (1-p_D)^2$ instead of $\sim (1-p_D)$) in the limit of large system sizes $N$. Then, in the presence of global depolarizing noise in the system, we can argue that the CFI is always strictly a lower bound to the true value of the quantum Fisher information with the fixed measurement setting considered here.

\begin{figure}
    \centering
\includegraphics[width=\linewidth]{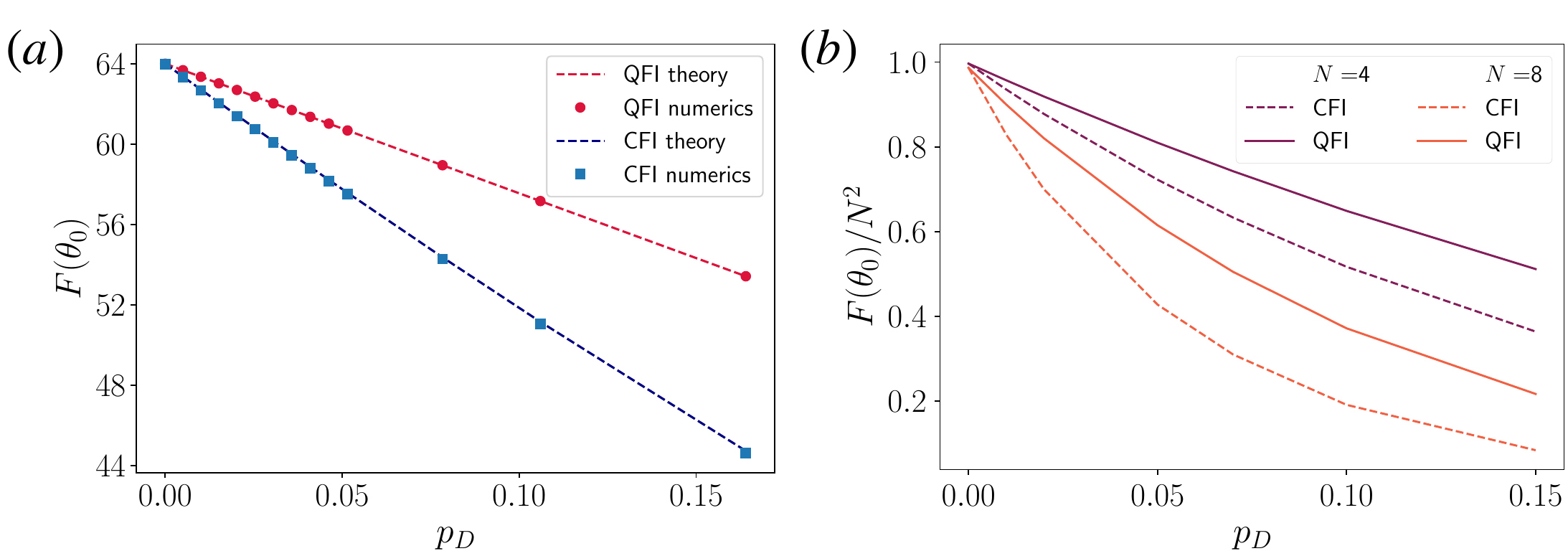}
    \caption{$(a)$ QFI and CFI of the GHZ state for $N=8$ qubits, in the presence of global depolarization of stength $p_D$. The points represent the numerical results while the dashed lines correspond to Eq.~\eqref{eq:qfidepo} and Eq.~\eqref{eq:CFIdepo}. $(b)$ QFI and CFI of the GHZ state simulated using 'qasm\_simulator'. Noise, in the form of a depolarizing channel, is added to each CNOT gate necessary to prepare the GHZ state. Its strength is labeled by $p_D$.}
    \label{fig:CFIdepo}
\end{figure}

In Fig.~\ref{fig:CFIdepo} we plot numerical results for comparing the QFI and the CFI in the presence of depolarization errors. We consider two cases, in Fig.~\ref{fig:CFIdepo}$(a)$ we calculate those quantities on the state defined as
\begin{equation}
    \rho=(1-p_D)\ket{\psi}\bra{\psi}+p_D\frac{\mathbb{1}}{d},
\end{equation}
where $d$ is the Hilbert space dimension ($d=2^N)$. For the state $\rho$ we have analytical predictions (dashed lines according to Eq.~\eqref{eq:qfidepoMT} and Eq.~\eqref{eq:CFIdepo}) that can be compared with numerical results according to the Hellinger method for estimating the CFI~\cite{Pezze_2016_PNAS}. We observe that the CFI decreases faster than QFI, enforcing our statement that it is important to find a direct and reliable estimator of the QFI.
In Fig.~\ref{fig:CFIdepo}$(b)$ we prepare the GHZ state on the simulator `qasm\_simulator'~\cite{IBMQ_ref}, by means of a Hadamard gate and $N$-1 CNOT gates. On each CNOT we add depolarizing noise whose strength $p_D$ we can tune.
We observe that the CFI and QFI both decrease, as expected when increasing $p_D$ and the number of qubits in the system. Also, as in the previous case, the CFI is always strictly smaller than the QFI for any value of $p_D>0$.

\subsection{$\mathcal{G}_j$ as a function of the number of unitaries and readout error}\label{SM:choiceofnu}

In this section, we study numerically the estimator $\hat{\mathcal{G}}_j$ in Eq.~\eqref{eq:estimatorMT}. We employ the IBM quantum simulator for providing an estimate of the scaling of $\hat{\mathcal{G}}_j$ as a function of the number of unitaries $N_U$ in the randomized measurement protocol in the calibration step. 
We induce noise in the circuit as a readout error $p_{\textrm{meas}}$ according to the noise model employed in Appendix~\ref{app:originnoise}.
In Fig.~\ref{fig11} we plot $\hat{\mathcal{G}}_j$ for different values of $p_{\mathrm{meas}}$, as a function of $N_U$. The estimation is compatible with the theoretical values (dashed lines) within error bars, for any value of $p_{\mathrm{meas}}$. We observe the error bars on the estimation decreases with $N_U$, for fixed $N_M=1000$. For the value of $N_U$ used in our experimental protocol ($N_U\sim 200$)  we observe an uncertainty of $\sim 1\%$ on the estimation of $\hat{\mathcal{G}}_j$. 
Increasing the number of unitaries used does not improve the estimation significantly. Hence, we choose $N_U=200$.

\begin{figure}[htbp!]
    \centering
    \includegraphics[width=0.5\linewidth]{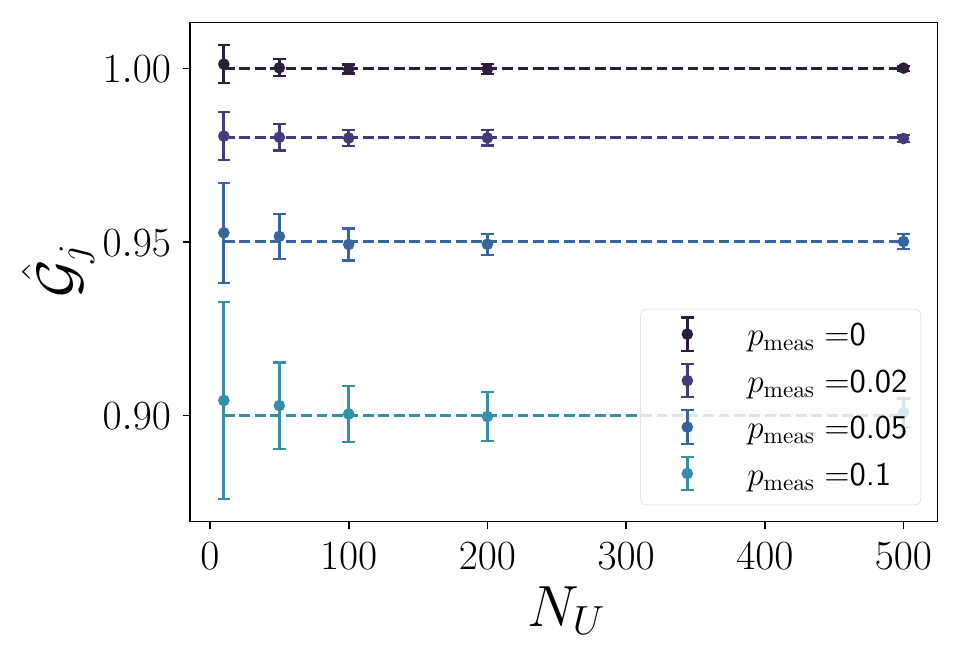}
    \caption{Numerical simulation of the calibration protocol on the IBMQ quantum simulator for a two qubit system. We plot $\hat{\mathcal{G}}_j$ as a function of $N_U$, for $N_M=1000$ and varying the readout error $p_{\mathrm{meas}}$.}
    \label{fig11}
\end{figure}

\subsection{Scaling of the measurement budget for the lower bound $F_2$}\label{SM:measbudget}
\begin{figure}
    \centering
    \includegraphics[width=0.5\linewidth]{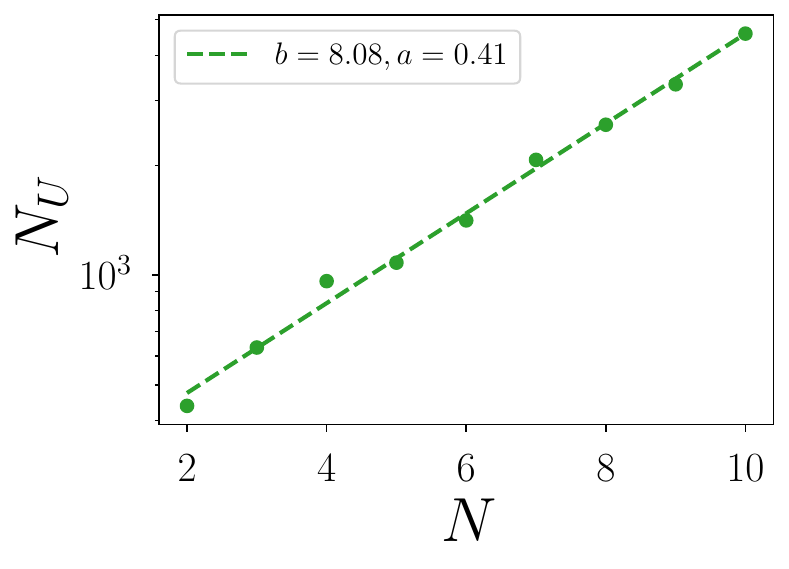}
    \caption{Numerical simulations to provide the number unitaries $N_U$ required to estimate $F_2$ below an error of $\mathcal{E} = 10\%$ for a GHZ state with respect to $A = \frac{1}{2} \sum_{j = 1}^{N} \sigma_j^{z}$. We simulate $N_M=1000$ computational basis measurements per unitary.  The dashed line is an exponential fit of the type $2^{b +aN}$ highlighting the scaling as a function of the system size $N$.}
    \label{fig12}
\end{figure}

In Fig.~\ref{fig12}, we provide numerical simulations to extract the scalings of the statistical errors on our highest measured lower bounds $F_2$. We consider an $N$-qubit pure GHZ state and consider once again the Hermitian operator $A = \frac{1}{2} \sum_{j = 1}^{N} \sigma_j^z$. We simulate the protocol by applying $N_U$ local random unitaries $U^{(r)}$ with $r = 1, \dots , N_U$ with $N_M = 1000$ projective computational basis measurements per unitary to obtain batch estimates $\hat{F}_2$ using $N_B = 10$ batches. 
The estimation is realized using Eq.~\eqref{eq:batchestimators} of the main text, however we do not consider here common randomized measurement, i.e. we take $\sigma=0$ in Eq.~\eqref{eq:shiftedshadow}.
The average statistical error $\mathcal{E}$ is calculated by averaging the relative error ${\mathcal{E}} = |\hat{F}_2 - F_2|/F_2$ over 100 numerically simulated experimental runs for different values of $N_U$. We find the maximum value of $N_U$ for which we obtain $\mathcal{E} \leq 0.1$ for different system sizes $N$ by using a linear interpolation function.

\subsection{Numerical simulation of the experiment for perfect GHZ states and readout errors}
\begin{figure}
    \centering
    \includegraphics[width=0.5\linewidth]{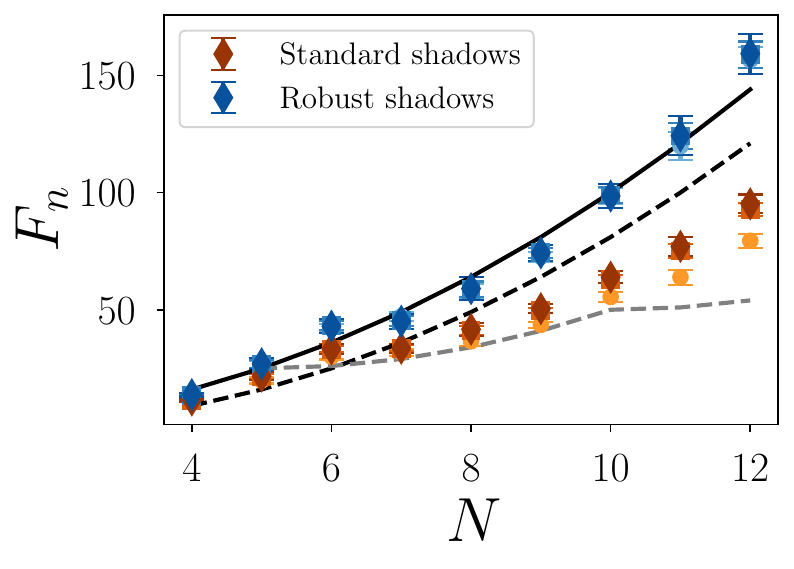}
    \caption{Numerical simulation of the experimental procedure, for perfect GHZ states but including readout errors. 
    As also mentioned in the main text, this figure shows $F_0$, $F_1$, $F_2$ (light to dark with circle, square and diamond respectively) as a function of the number of qubits $N$, where we fix as always the operator $A = \frac{1}{2} \sum_{j = 1}^{N} \sigma_j^{z}$. The solid line is the exact value of the QFI $F_Q = N^2$ for pure GHZ states. The dashed black line corresponds to the entanglement witness \mbox{$\Gamma (N,k=N-1) = (N-1)^2$}.
    The dashed grey line corresponds to the entanglement witness $\Gamma (N,k=5)$.}
    \label{fig13}
\end{figure}

In this last section, we provide the measurement of the lower bounds via a classical numerical experiment for GHZ states prepared without any state preparation errors (perfect GHZ states).
We take the same measurement budget as applied in the experimental procedure (c.f Sec.~\ref{sec:budget} of the main text).
Here again,  we consider $\sigma=0$ in Eq.~\eqref{eq:shiftedshadow}.
Additionally, we consider that the single qubit random unitary operations are done perfectly and take into account only readout errors with a probability of $p_{\rm meas} = 1.4\%$ as recorded for the IBM superconducting qubit device `ibm\_prague'~\cite{IBMQ_ref}. The results are shown in Fig.~\ref{fig13}.

\end{document}